\documentclass[traditabstract]{aa}  
\usepackage{graphicx}
\usepackage{txfonts}
\usepackage{hyperref}
\newcommand\fnurl[2]{%
  \href{#2}{#1}\footnote{\url{#2}}%
}
\begin{document}
   \title{Mass assembly in quiescent and star-forming galaxies since
     $z \simeq 4$ from UltraVISTA \thanks{Based on data products from
       observations made with ESO Telescopes at the La Silla Paranal
       Observatory under ESO programme ID 179.A-2005 and on data
       products produced by TERAPIX and the Cambridge Astronomy Survey
       Unit on behalf of the UltraVISTA consortium. }\fnmsep{}\thanks{
       Catalogues are only available in electronic form at the CDS via
       anonymous ftp to cdsarc.u-strasbg.fr (130.79.128.5) or via
       \url{http://cdsweb.u-strasbg.fr/cgi-bin/qcat?J/A+A/}}}

   \author{
O. Ilbert \inst{1}
\and H.~J.~McCracken \inst{2}
\and O.~Le F\`evre \inst{1}
\and P.~Capak \inst{3}
\and J.~Dunlop \inst{4}
\and A.~Karim \inst{5,6}
\and M.~A. Renzini \inst{7}
\and K.~Caputi \inst{8}
\and S.~Boissier \inst{1}
\and S.~Arnouts \inst{9}
\and H.~Aussel \inst{10}
\and J.~Comparat \inst{1}
\and Q.~Guo \inst{11}
\and P.~Hudelot \inst{2}
\and J.~Kartaltepe \inst{12}
\and J.~P.~Kneib \inst{1}
\and J.~K.~Krogager \inst{13}
\and E.~Le Floc'h \inst{14}
\and S.~Lilly \inst{15}
\and Y.~Mellier \inst{2}
\and B.~Milvang-Jensen \inst{13}
\and T.~Moutard \inst{1}
\and M.~Onodera \inst{15}
\and J.~Richard \inst{16}
\and M.~Salvato \inst{17,18}
\and D.~B. Sanders \inst{19}
\and N.~Scoville \inst{20}
\and J.D.~Silverman \inst{21}
\and Y.~Taniguchi \inst{22}
\and L.~Tasca \inst{1}
\and R.~Thomas \inst{1}
\and S.~Toft \inst{13}
\and L.~Tresse \inst{1}
\and D.~Vergani \inst{23}
\and M.~Wolk \inst{2}
\and A.~Zirm \inst{13}
         }
\institute{
Aix Marseille Universit\'e, CNRS, LAM (Laboratoire d'Astrophysique de Marseille) UMR 7326, 13388, Marseille, France \\   \email{olivier.ilbert@lam.fr}
\and
Institut d'Astrophysique de Paris, UMR7095 CNRS, Universit\'e Pierre et Marie Curie, 98 bis Boulevard Arago, 75014 Paris, France
\and
Spitzer Science Center, California Institute of Technology, Pasadena, CA 91125, USA
\and
Institute for Astronomy, University of Edinburgh, Royal Observatory, Edinburgh, EH9 3HJ, UK
\and
Institute for Computational Cosmology, Durham University, South Road, Durham, DH1 3LE, UK
\and
Argelander-Institute of Astronomy, Bonn University, Auf dem H\"ugel 71, D-53121 Bonn, Germany 
\and
Dipartimento di Astronomia, Universita di Padova, vicolo dell'Osservatorio 2, I-35122 Padua, Italy
\and
Kapteyn Astronomical Institute, University of Groningen, P.O. Box 800, 9700 AV Groningen, the Netherlands
\and
Canada France Hawaii telescope corporation, 65-1238 Mamalahoa Hwy, Kamuela, Hawaii 96743, USA
\and
AIM Unit\'e Mixte de Recherche CEA  CNRS Universit\'e Paris VII UMR n158, France
\and
Partner Group of the Max-Planck-Institut f\"ur Astrophysik, National Astronomical Observatories, Chinese Academy of Sciences, Beijing 100012, China
\and
National Optical Astronomy Observatory, 950 North Cherry Avenue, Tucson, AZ 85719, USA
\and
Dark Cosmology Centre, Niels Bohr Institute, University of Copenhagen, Juliane Mariesvej 30, DK-2100 Copenhagen, Denmark
\and
Laboratoire AIM, CEA/DSM/IRFU, CNRS, Universit\'e Paris-Diderot, 91190 Gif, France
\and
Department of Physics, ETH Zurich, CH-8093 Zurich, Switzerland
\and
CRAL, Universit\'e Lyon-1 and CNRS-UMR 5574, 9 avenue Charles Andr\'e, 69561 Saint-Genis Laval Cedex, France
\and
Max-Planck-Institut f\"ur Extraterrestrische Physik, Postfach 1312, D-85741, Garching bei M\"unchen, Germany
\and
Excellence Cluster, Boltzmann Strasse 2, 85748 Garching , Germany
\and
Institute for Astronomy, 2680 Woodlawn Dr., University of Hawaii, Honolulu, Hawaii, 96822
\and
California Institute of Technology, MC 105-24, 1200 East California Boulevard, Pasadena, CA 91125
\and 
Kavli Institute for the Physics and Mathematics of the Universe, Todai Institutes for Advanced Study, the University of Tokyo, Kashiwa, Japan 277-8583 (Kavli IPMU, WPI)
\and
Research Center for Space and Cosmic Evolution, Ehime University, Bunkyo-cho 2-5, Matsuyama 790-8577, Japan
\and 
INAF - IASFBO, via P. Gobetti 101, 40129, Bologna, Italy 
}

   \date{Received ... ; accepted ...}

   \abstract{We estimate the galaxy stellar mass function and stellar
     mass density for star-forming and quiescent galaxies with
     $0.2<z<4$. We construct a large, deep ($K_{\rm s}<24$) sample of
     220,000 galaxies selected using the new UltraVISTA DR1 data
     release. Our analysis is based on precise 30-band photometric
     redshifts. By comparing these photometric redshifts with 10,800
     spectroscopic redshifts from the zCOSMOS bright and faint
     surveys, we find a precision of $\sigma_{\Delta z / (1+z)}=0.008$
     at $i^+<22.5$ and $\sigma_{\Delta z / (1+z)}=0.03$ at $1.5<z<4$.
     We derive the stellar mass function and correct for the Eddington
     bias.  We find a mass-dependent evolution of the global and
     star-forming populations, with the low-mass end of the mass
     functions evolving more rapidly than the high-mass end. This
     mass-dependent evolution is a direct consequence of the star
     formation being ``quenched'' in galaxies more massive than ${\cal
       M} \gtrsim 10^{10.7-10.9} {\cal M}_{\sun}$. For the mass
     function of the quiescent galaxies, we do not find any
     significant evolution of the high-mass end at $z<1$; however we
     observe a clear flattening of the faint-end slope. From $z\sim3$
     to $z\sim1$, the density of quiescent galaxies increases over the
     entire mass range. Their comoving stellar mass density increases
     by 1.6 dex between $z \sim 3$ and $z \sim 1$ and by less than 0.2
     dex at $z<1$. We infer the star formation history from the mass
     density evolution. This inferred star formation history is in
     excellent agreement with instantaneous star formation rate
     measurements at $z<1.5$, while we find differences of 0.2 dex at
     $z>1.5$ consistent with the expected uncertainties. We also
     develop a new method to infer the specific star formation rate
     from the mass function of star-forming galaxies. We find that the
     specific star formation rate of $10^{10-10.5} {\cal M}_{\sun}$
     galaxies increases continuously in the redshift range $1<z<4$.
     Finally, we compare our results with a semi-analytical model and
     find that these models overestimate the density of low mass
     quiescent galaxies by an order of magnitude, while the density of
     low-mass star-forming galaxies is successfully reproduced.
     }

  \maketitle
%

\section{Introduction}

The galaxy stellar mass function (hereafter MF) is a fundamental
indicator of the physical processes that regulate mass assembly in
galaxies across cosmic time. Stellar mass assembly in galaxies is
believed to result from several physical processes such as star
formation from in-situ or accreted gas as well as from major or minor
mergers. The star formation may also be stopped, or ``quenched'', by
external process such as ``feedback'' in Active Galactic Nuclei (AGN)
for galaxies residing in more massive dark matter haloes, or by other
processes such as supernovae-driven winds in less massive haloes. The
relative contribution and operating timescales of these different
processes are still a matter of debate.

The evolutionary tracks of the MF as a function of look-back time
reveal the major paths taken by different galaxy populations across
cosmic time. Despite the considerable progress made over the last
decade, our understanding of the early evolution of stellar mass
growth is still incomplete and several puzzling issues have yet to be
resolved: {\it (i)} The stellar mass density of the most massive
galaxies ($10^{11.7} {\cal M}_{\sun}$) at $z\sim 3$ is almost
identical to the local measurement (Perez-Gonzalez et al. 2008). This
result requires a powerful feedback mechanism in order to halt star
formation in massive galaxies, like AGN feedback (e.g. Bower et
al. 2006, Croton et al. 2006). However, numerous systematic
uncertainties could still be hiding an evolution of the high-mass end
(Marchesini et al. 2009). In particular, if both random and systematic
errors are considered, Marchesini et al. (2009) ``can not exclude a
strong evolution (by as much as a factor of $\sim$50) in the number
density of the most massive galaxies ($>10^{11.5} {\cal M}_{\sun}$)
from z = 4.0 to z = 1.3''. {\it (ii)} The stellar mass density of the
most massive quiescent galaxies undergoes little evolution (less than
0.2 dex) from $z\sim1$ to $z\sim 0.1$ (Arnouts et al. 2007, Pozzetti
et al. 2010, Ilbert et al. 2010).  Therefore, major merger activity
must be relatively limited for this population since $z\sim1$, in
agreement with galaxy merger rate measurements (e.g. Lopez-Sanjuan et
al., 2012). {\it (iii)} The stellar mass density increases
continuously with cosmic time. The star formation rate (hereafter SFR) integrated
along cosmic time and the stellar mass density evolution should
provide a coherent picture (Arnouts et al. 2007, Boissier et
al. 2010), but are still difficult to reconcile unless the initial
mass function (IMF) changes with time, as advocated by several authors
(e.g. Wilkins et al. 2008, Lu et al. 2012). A significant uncertainty
remains on the contribution of low mass galaxies which could impact
the global stellar mass density at $z>2$ (Mortlock et al. 2011,
Santini et al. 2012). {\it (iv)} Red galaxies are created very
efficiently at $1<z<2$ (Cirasuolo et al. 2007, Arnouts et al. 2007,
Ilbert et al. 2010, Cassata et al. 2011). This implies that the
quenching of star forming galaxies must be extremely efficient at
$z>1$. Arnouts et al. (2007), Ilbert et al. (2010) and Kajisawa et
al. (2011) found that the stellar mass density of quiescent galaxies
increased by roughly one order of magnitude from $z\sim2$ to $z\sim1$
(2.5Gyr) as star-formation stops in galaxies which then migrate into
the ``red sequence'' of passively evolving galaxies. But the amount of
evolution is still debated. Brammer et al. (2011) find also an
evolution of the quiescent population at $1<z<2$, but at a smaller
rate of 0.5 dex for the massive population. Ilbert et al. (2010)
identify $z\sim1$ as a transition epoch in the assembly of the most
massive quiescent galaxies (the density of the most massive galaxies
ceases to increase at $z<1$). But from the Brammer et al. (2011)
study, no clear transition occurs at $z \sim 1$. Therefore, a key
point is to consolidate these measurements at $z<2$ and follow the
growth of the quiescent population to even earlier times.

Beyond $z\sim1$ observations are challenging, and are currently
limited either by depth or small numbers of objects in relatively
small fields, making the computation of the massive end of the MF
sensitive to cosmic variance and the lower mass end difficult to
constrain. The identification of a robust quiescent galaxy sample
requires accurate photometric redshifts with a low number of
catastrophic failures. A field of at least one square degree or more
is necessary to provide a large volume and minimise cosmic
variance. Moreover, samples need to be deep enough to probe the
low--mass end of the MF beyond $z\sim2$. The availability of deep wide
field near-IR multi-band photometry is essential. Spectral features
like the D4000 or Balmer break move into the near-IR at $z>1.5$, and
several near-IR bands are required to properly sample the spectral
energy distribution (SED) and enable a stable photometric redshift and
type classification from SED-fitting techniques.

The COSMOS field is one of the best available fields to derive the MF
thanks to the large area (2 deg$^2$) and the large amount of deep
($I_{AB}\sim 26.5$) multi-wavelength data available (more than 35
bands). Several papers showing the MF evolution in the COSMOS field
have been published (Drory et al. 2009, Ilbert et al. 2010, Pozzetti
et al. 2010, Dominguez-Sanchez et al. 2011). New photometric and
spectroscopic datasets have been obtained in the last two years which
allow us to greatly reduce the systematic uncertainties in the MF
estimate at $z>1$.  The first UltraVISTA DR1 data
release\footnote{${\rm
    www.eso.org/sci/observing/phase3/data\_releases/ultravista\_dr1.html}$}
(McCracken et al. 2012) covers $1.5~\mathrm{deg}^2$ in four
near-infrared filters $Y$, $J$, $H$ and $K_{\rm s}$. The DR1 data are
at least one magnitude deeper than previous COSMOS near-infrared
datasets (McCracken et al. 2010) and also provide new Y-band
photometric information. Since the Balmer break lies between
UltraVISTA filters at $z>1.3$, we can now derive photometric redshifts
which are more robust at high redshift.  Secondly, almost 35,000 new
spectra are now available in the COSMOS field. This sample includes
9,900 spectroscopic redshifts at $z>1$, including extremely faint
objects. Such a spectroscopic sample is essential to ensure that our
analysis can be extended at high redshift. New NIR spectroscopic
samples (WFMOS and MOIRCS on Subaru and WFC3/HST grism data) contain
spectroscopic redshifts for the quiescent and the dusty populations at
$z>1.5$. Therefore, we can ensure that our photo-z are robust at $z>1$
and we are now able to extend MF measurements to $z=4$ and to lower
stellar masses than previous studies in the same field.

In this paper, we extend MF and stellar mass density measurements out
to $z=4$ and for stellar masses down to $10^{10.3} {\cal M}_{\sun}$
for the global population. The new data sets are introduced in
\S\ref{Data}.  The photometric redshifts and associated physical
properties are discussed in \S\ref{phy}. The method used to estimate
the MF and the associated uncertainties is given in
\S\ref{estimate}. We present the measured MF and stellar mass density
for the full, star-forming and quiescent galaxy samples in
\S\ref{MF}. Results are discussed and compared to semi-analytical
models in \S\ref{discussion}. In particular, we investigate systematic
uncertainties linked to the choice of the star formation histories in
our models.
Throughout this paper, we use the standard cosmology
($\Omega_m~=~0.3$, $\Omega_\Lambda~=~0.7$ with
$H_{\rm0}~=~70$~km~s$^{-1}$~Mpc$^{-1}$). Magnitudes are given in the
$AB$ system (Oke 1974). The stellar masses are given in units of solar
masses (${\cal M}_{\sun}$) for a Chabrier (2003) IMF.

\section{Data description}\label{Data}

\subsection{Preparation of stacked images  and confidence maps}
\label{sec:prep-image-stacks}

Our photometric catalogue comprises near-infrared data taken with the
VIRCAM (Emerson \& Sutherland 2010) on the VISTA telescope as part of
the UltraVISTA project and optical broad and intermediate-band data
taken with the SUPRIME camera on Subaru in support of the COSMOS
project (Capak et al. 2007). The near-infrared data we use here
corresponds to the UltraVISTA DR1 data release fully described in
McCracken et al. (2012).

To construct our multi-band catalogue, we first downloaded all COSMOS
tiles and confidence maps from the
\fnurl{IRSA}{http://irsa.ipac.caltech.edu/Missions/cosmos.html} COSMOS
archive, choosing where possible the ``best'' seeing images. Since the
IRSA tiles have the same tangent point and pixel scale as UltraVISTA
DR1, they may be simply ``pasted together'' (i.e., \textit{without}
image resampling) using the \texttt{swarp} software to produce a
single, large image which is pixel-matched to UltraVISTA
DR1. Catalogues can then be simply generated using \texttt{SExtractor}
(Bertin \& Arnouts 1996) in ``dual-image'' mode, using matched
apertures on each image. The list of broad and intermediate band /
narrow band images used is as follows: $u^*$, $B_J$, $V_J$, $r^+$,
$i^+$, $z+$, $IA484$, $IA527$, $IA624$, $IA679$, $IA738$, $IA767$,
$IA427$, $IA464$, $IA505$, $IA574$, $IA709$, $IA827$, $NB711$,
$NB816$. Table 1. in Ilbert et al (2009) lists the effective
wavelengths of each of these filters (note that we do not use the
Subaru $g$-band data as this has particularly poor seeing).  All four
UltraVISTA bands are used: $Y$, $J$, $H$, $K_{\rm s}$; the depths for
the UltraVISTA DR1 are given in Table 1 of McCracken et al. (2012).

Before catalogue extraction, we must first take into account the large
variation in seeing between different COSMOS images and also within
the UltraVISTA stacks themselves. As already documented in McCracken
et al. (2012), in the UltraVISTA $H$ and $K_{\rm s}$ stacks there are
``columns'' of different seeing as a consequence of the different
VISTA pawprints contained in the final stacks being taken under
different seeing conditions. To correct for this, we first construct
for each $H$ and $K_{\rm s}$ stack six separate stacks comprising the
six individual VISTA ``pawprints''.  We then measure the average
seeing on each of six pawprints using the \texttt{PSFex} software
(Bertin et al. 2011), which corresponds to a fit of a Moffat (1969)
profile. Next, the individual pawprints are convolved by a Gaussian
profile calculated to bring the PSF on the final images to
$1.1\arcsec$. A similar procedure is adopted to homogenise the COSMOS
broad and intermediate-band images: the seeing is measured on each
stack and the images are then convolved with a Gaussian profile to
bring the final image to $1.1\arcsec$ which is the worst PSF among all
bands (the $Y$ band from ultraVISTA).

\subsection{IRAC observations}
\label{sec:irac-observations}
The IRAC data consist of all cryogenic data (Sanders et al. 2007) and
the data from the Warm mission SEDS program (Fazio et al. 2012)
covering $\sim 0.1 \mathrm{deg}^2$ in the center of the field.  The
data were reduced with the
\fnurl{MOPEX}{http://ssc.spitzer.caltech.edu/postbcd/} software
package. Photometric measurements were made using \texttt{SExtractor}
in dual image mode with the Subaru $i$-band image as a detection image
and the IRAC image for measurement. This improves photometric accuracy
when optical sources are close to each other by making use of the
\texttt{SExtractor} aperture de-blending routines. As in described in
Sanders et al. (2007) photometric measurements were made in a
3.8\arcsec diameter aperture and corrected to pseudo-total using a
statistical aperture correction. We adopt a 3.8\arcsec aperture
  which is a good trade-off between a small aperture which limits the
  noise created by blending, and a large aperture which reduces the
  correction needed to estimate the total fluxes of bright galaxies.
A comparison between these IRAC fluxes in the $i$- and IRAC selected
catalogues (Ilbert et al. 2010) shows only a small systematic offset
which is to be expected as a consequence of the update to the IRAC
calibration.

\subsection{Source catalogue extraction and merged catalogue creation}
\label{sec:prep-detect-imag}

In extracting our source catalogues using \texttt{SEXtractor}, the
choice of the detection image is important. Our scientific objectives
drive us to use the longest possible wavelengths in order to reliably
detect galaxies at intermediate to high redshifts. On the other hand,
we also want a sample which is as \textit{complete} as possible to the
faintest possible limits in stellar mass. We construct a detection
image using \texttt{SWarp} (Bertin et al. 2002) from the chi-squared
sum of the (non-convolved) UltraVISTA DR1 $YJHK_{\rm s}$ images,
following the techniques outlined in Szalay et al. (1999). This
ensures that all sources detected in at least one VISTA band are
included in the final catalog.

With this detection image and a set of PSF-homogenised images in hand,
we can proceed to catalogue extraction. In order to make reliable
magnitude error estimates on stacked, convolved data which has a
non-negligible amount of correlated noise we use for each Subaru band
pre-computed effective gain values (Capak et al. 2007) in combination
with the \texttt{MAP\_RMS} confidence maps supplied by IRSA. This is
particularly important for the shorter wavelength data which saturates
at relatively bright magnitudes. For the UltraVISTA data, where we use
weight maps as opposed to RMS-maps we set the \texttt{SEXtractor}
option \texttt{RESCALE\_WEIGHTS} to \texttt{N} to ensure that image
convolution has no effect on image noise measurement and that the
weight-maps are not rescaled automatically as is normally the case in
\texttt{SExtractor}. To account for additional noise sources not
accounted for in the RMS maps such as errors in the background
subtraction, zero-point errors, confusion, morphological aperture
corrections, and other effects the errors were multiplied by a factor
of 1.5 to match the measured noise on the extracted photometry.

For each detected source we measure aperture magnitudes into a
$3\arcsec$ diameter circle, and ``pseudo-total'' ``Kron'' magnitudes
(Kron, 1980), corresponding to \texttt{SExtractor}'s
\texttt{MAG\_APER} and \texttt{MAG\_AUTO} respectively. Next,
catalogues from each band were merged together into a single
\texttt{FITS} table and galactic extinction values computed at each
object position using the Schlegel et al. (1998) dust maps were
added. In addition object mask flags indicating bad regions in optical
and near-infrared bands were included, and saturated pixels in the
optical bands were flagged by using the appropriate
\texttt{FLAG\_MAP}s at the extraction stage.

In each band, for the purposes of photometric redshift measurements,
aperture magnitudes are corrected to pseudo-total magnitudes using a
band-dependent aperture correction computed using a stellar
curve-of-growth method (although the images are PSF-homogenised, there
are still small residual PSF variations band-to-band). The SED fitting
is performed on these magnitudes. However, this aperture correction is
not enough to capture the total flux of large and bright galaxies. As
a consequence, we would underestimate the stellar masses of these
galaxies. To get the correct stellar masses, we compute an average
transformation between these aperture corrected magnitudes and
\texttt{SExtractor}'s \texttt{MAG\_AUTO} over all broad-band filters
for each object and then apply this adjustment to the stellar masses.

The $i-$ selected catalogue from Capak et al. (2007) was
cross-matched to the near-infrared UltraVISTA catalogue. This $i-$
selected catalogue contains GALEX UV magnitudes computed by Zamoski et
al (2007). GALEX fluxes were measured using PSF fitting method.

This $i-$ selected catalogue contains also the IRAC fluxes as
  described above. Some red galaxies could be too faint to be detected
  in the $i$-band selected catalogue (e.g. quiescent galaxies at
  $z>2$). Nevertheless, such galaxies could be detected in the IRAC
  images. For these sources, we include the 3.8'' aperture fluxes from the IRAC
  selected catalogue of Ilbert et al. (2010).  

Finally, a significant fraction of the objects (2\%) were flagged
  as power-law sources using the criteria established by Donley et
  al. (2012). Since in these cases the IRAC flux could be dominated by
  AGN emission, we do not use the IRAC flux for this population.

\begin{figure}
\centering \includegraphics[width=9cm]{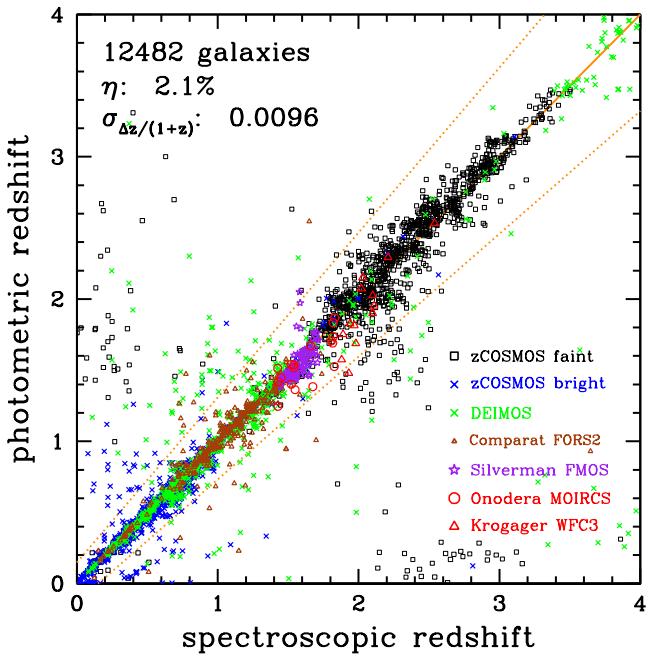}
\caption{Photometric redshifts versus spectroscopic redshifts. Only
  secure spectroscopic redshifts at $K_{\rm s}<24$ are
  considered. Different symbols correspond to the spectroscopic
  samples labeled in the right.}
           \label{zp_zs}%
\end{figure}

\begin{table}[htb!]
\begin{center}
\begin{tabular}{l c c c c c c } \hline

    spectroscopic       &           Nb spec-z  & $z_{med}$  &  $I_{med}$  &   $\sigma_{\Delta z/(1+z)}$ &  $\eta$(\%)  \\
    survey              &        $K_{\rm s}<24$ &           &           &                     &        \vspace{0.2cm} \\ \hline

    zCOSMOS bright      &             9389     &  0.50     &   21.4    &      0.0080         &  0.6      \\ 
    Kartaltepe 2013    &              570     &  0.73     &   22.0    &      0.0105         &  3.2     \\ 
    Comparat 2013       &              382     &  1.12     &   22.6    &      0.0163         &  4.7      \\ 
    Capak 2013          &              631     &  1.15     &   23.5    &      0.0213         &  9.5     \\ 
    Onodera 2012        &               17     &  1.55     &   23.9    &      0.0446         &  0.0      \\ 
    Silverman 2013      &               88     &  1.58     &   23.2    &      0.0259         &  1.1      \\ 
    Krogager 2013       &               13     &  2.02     &   24.8    &      0.0708         &  7.7      \\ 
    zCOSMOS faint       &             1392     &  2.15     &   23.6    &      0.0297         &  7.7      \\ 
\hline
\end{tabular}
\caption{Characteristics of the spectroscopic redshift samples and
  photometric redshift accuracy. Only the most secure spectroscopic
  redshifts at $K_{\rm s}<24$ are considered. The median redshift and
  magnitude are provided for each sample. \label{zs}}
\end{center}
\end{table}

\section{Photometric redshifts and physical parameters}
\label{phy}

\subsection{Photometric redshift estimation}
\label{sec:phot-redsh-estim}

We derive our photometric redshifts following the recipes outlined in
Ilbert et al. (2009). The photometric redshifts are derived using ``Le
Phare'' (Arnouts et al. 2002, Ilbert et al. 2006). Ilbert et
al. (2009) used 31 templates including elliptical and spiral galaxies
from the Polletta et al. (2006) library and 12 templates of young and
blue star-forming galaxies generated with Bruzual and Charlot stellar
population synthesis models (2003, hereafter BC03). Using a
spectroscopic sample of quiescent galaxies, Onodera et al. (2012)
showed that the estimate of the photo-z for the quiescent galaxies in
Ilbert et al. (2009) were underestimated at $1.5<z<2$. We improve
the photo-z for this specific population by adding two new templates
of elliptical galaxies generated with BC03. These two new templates
are generated assuming an exponentially declining SFR
with a short timescale $\tau=0.3Gyr$ and two different metallicities
($Z=0.008$ and $Z=0.02$ i.e. $Z{\sun}$). They include 22 ages well
sampled between 0.5Gyr and 4Gyr old. Thanks to a better sampling of
younger ages than our previous library, they improve the photo-z for
the quiescent population at $z > 1.5$.

Extinction is added as a free parameter (E(B-V)$<$0.5) and several
extinction laws are considered (Calzetti et al. 2000, Prevot 1984 and
a modified version of the Calzetti laws including a bump at
2175\AA). We do not add any extinction for the templates redder than
S0. We also do not allow additional extinction for the templates
redder than Sc (the Sa, Sb, Sc templates from Polletta et al. 2007
already include some extinction). Emission lines are added to the
templates using an empirical relation between the UV light and the
emission line fluxes (Ilbert et al. 2009). By contrast with Ilbert et
al. (2009), we assign the redshift using the median of the
marginalised probability distribution function rather than the minimum
$\chi^2$. While the results are broadly similar between the two
methods, using the median has the advantage of producing more reliable
error bars for the photometric redshifts\footnote{We integrate the PDF
  over 68\% of its area around the median solution} and reducing the
effect of aliasing in the photometric redshift space.

We combine several spectroscopic samples to test the accuracy of the
photometric redshifts, including zCOSMOS-bright with 20700 bright
VIMOS/VLT spectra selected at $i^+<22.5$ (Lilly et al. 2007), zCOSMOS
faint with 9500 faint VIMOS/VLT spectra selected at $1.5<z<3$ (Lilly
et al., in preparation), 2300 DEIMOS/Keck redshifts which combined
several selected sub-populations of blue star-forming and infrared
galaxies at $0.5<z<6$ (Kartaltepe et al., in preparation, Capak et
al., in preparation), 835 FORS2/VLT redshifts at $0.6<z<1.8$ (Comparat
et al., in preparation), 138 FMOS/Subaru redshifts at $1.4<z<1.8$
(Silverman et al., in preparation), 18 faint quiescent galaxies at
$z<1.9$ obtained with MOIRCS/Subaru (Onodera et al. 2012) and 16 faint
quiescent galaxies at $1.85<z<2.6$ obtained with the WFC3 grism
observations from the 3D-HST survey (Krogager et al., in
preparation). We keep only the most secure spectroscopic redshifts
(e.g. flag 3 and 4 for zCOSMOS) at $K_{\rm s}<24$ which reduces to
12482 the number of spectroscopic redshifts used for the
comparison. The comparison between photometric and spectroscopic
redshifts is shown in Figure.\ref{zp_zs} and the accuracy obtained for
each spectroscopic sample is listed in Table~1. The fraction of
  catastrophic failures $\eta$ is defined as
  $|z_{phot}-z_{spec}|/(1+z_{spec}) >0.15$. The redshift accuracy
  $\sigma_{\Delta z /(1+z_{\rm s})}$ is computed using the normalised
  median absolute deviation (Ilbert et al. 2006).

Figure.\ref{zp_zs} shows that the photometric redshift accuracy has
two regimes. At $z<1.5$, the Balmer break falls between the
intermediate band filters. In this regime, the spectroscopic sample is
dominated by the zCOSMOS bright sample selected at $i^+<22.5$ (blue
crosses). The precision is better than $1\%$ with less than 1\% of
catastrophic failures. Even as faint as $i^+ \sim 24$, comparisons
with the DEIMOS sample show a better than $3\%$ precision at $z<1.5$
(green crosses). At intermediate redshifts $1.4<z<2$, the FMOS sample
from Silverman et al. shows an excellent precision of 0.026. However
we note that $H_\alpha$ is identified in the NIR spectra from a
knowledge of photo-z which could explain the low failure rate. At
higher redshifts $1.5<z<4$ the accuracy of the photo-z has been tested
against the zCOSMOS faint sample and faint DEIMOS spectra. Here the
precision is around 3\%, showing that we can safely extend our
analysis at $z>1.5$ for faint star-forming galaxies.  Thanks to new
NIR spectroscopic samples, we can now test the accuracy of the photo-z
for the quiescent population at $z>1.5$. We combine the quiescent
sample of Onodera et al. (2012) and Krogager et al. (in
preparation). These galaxies are extremely faint with a median $i^+$
band magnitude of 24.5 and a median redshift of 1.8.  We have only one
outlier in 30 galaxies. The precision is as good as $\sigma_{\Delta z
  /(1+z_{\rm s})}=0.056$. We still detect a small bias with
$median(zp-zs)=-0.1$, which does not have significant impact on our
work.

We also compute the best-fitting $\chi^2$ using the stellar
library. The sources are classified as stars when the $\chi^2$
obtained with the star library is lower than the one obtained with the
galaxy library. Since near-infrared data are available, such criterion
is effective in isolating the stars.

   \begin{figure}
   \centering
   \includegraphics[width=9cm]{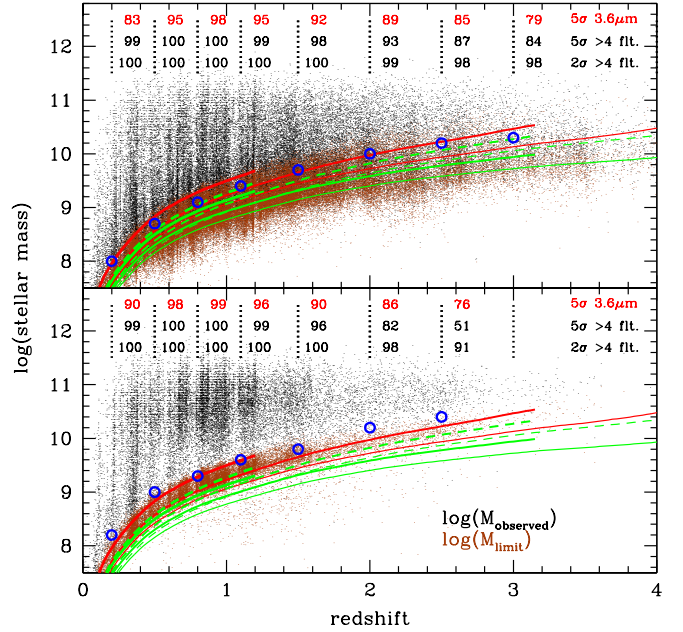}
   \caption{Stellar masses versus redshift for the $K_{AB}<24$
     selected sample. The top and bottom panels correspond to the full
     sample and the quiescent population, respectively. Black points
     correspond to the measured stellar masses. The brown points
     correspond to ${\cal M}_{limit}$ which is the lowest mass which
     could be observed for a given galaxy for a limit at
     $K_{AB}=24$. The blue circles correspond to the completeness
     limit chosen in this study. The red and green curves correspond
     to the two extreme templates with $\tau=0.1Gyr$ and $\tau=30Gyr$,
     respectively. We show 3 ages which are 0.9, 2 and 5 Gyr (from the
     bottom to the top, respectively). The dashed lines correspond to
     an extinction of $E(B-V)=0.2$. If the lines are not drawn, it
     means that either the age is older than the age of the Universe
     or that the condition $age/\tau>4$ implies a small
     extinction. We indicate at the top of each panel the
       percentage of galaxies detected at 3.6$\mu m$ (top line)
       and in at least four bands (detection thresholds at $>2\sigma$ and
       $>5\sigma$ for the middle and bottom lines, respectively).}
              \label{completeness}%
    \end{figure}

\subsection{Stellar masses and associated completeness}\label{stellarmasses}

We turn now to the estimation of physical properties based on these
photometric redshifts, first estimating galaxy stellar masses. We rely
on a model to convert the galaxy luminosity into stellar mass. We
generate a library of synthetic spectra normalised at one solar
mass. These synthetic spectra are fitted to the multi-colour
photometry described above using {\it Le\_Phare}. The physical
parameter called ``stellar mass'' in this paper corresponds to the
median of the stellar mass probability distribution marginalized over
all other parameters. The library of synthetic spectra is generated
using the Stellar Population Synthesis (SPS) model of Bruzual and
Charlot (2003). Several other models are available in the literature
(e.g. Fioc \& Rocca-Volmerange 1997, Maraston 2005, Bruzual 2007,
Conroy et al. 2009). The stellar masses could vary by 0.1-0.15 dex
depending on the considered SPS model (e.g. Walcher et al. 2011). For
consistency, we use the same default library as Ilbert et al. (2010):
we assume the Calzetti (2000) extinction law; emission line
contributions are included using an empirical relation between the UV
light and the emission line fluxes (Ilbert et al. 2009); we used three
different metallicities ($Z=0.004$, $Z=0.008$, $Z=0.02$
i.e. $Z{\sun}$); the star formation history declines exponentially
following $\tau^{-1} e^{-t/\tau}$. We considered nine possible $\tau$
values ranging from 0.1 Gyr to 30 Gyr. We will test the systematic
uncertainties linked to our choice of star formation history in
\S\ref{sfrhist}. Following Fontana et al. (2006), Pozzetti et
al. (2007), Ilbert et al. (2010), we impose the prior $E(B-V)<0.15$ if
$age/\tau>4$ (a low extinction is imposed for galaxies having a low
SFR).

Figure \ref{completeness} shows the stellar masses as a function of
redshift. We can detect galaxies with masses as low as ${\cal M}\sim
10^{10}{\cal M}_{\sun}$ at $z=4$. The mass limit depends on the
mass-to-light ratio of the considered template.  Figure
\ref{completeness} shows the stellar mass limit that can be reached
for different templates. For clarity, we show only two extreme
templates in terms of star formation history ($\tau=0.1Gyr$ and
$\tau=30Gyr$), with solar metallicity and we select three ages. Oldest
galaxies have the highest stellar mass limit. We follow a procedure
similar to Pozzetti et al. (2010) to define the stellar mass
completeness limit. We base our estimate on the 90\% of the templates
which are the most often fitted. In practice, we compute the lowest
stellar mass which could be detected for a galaxy with $log({\cal
  M}_{limit})=log({\cal M})+0.4(K-24)$, given a sample selected at
$K<24$ (brown points). At a given redshift, the stellar mass
completeness limit corresponds to the mass with 90\% of the galaxies
having their ${\cal M}_{limit}$ below the stellar mass completeness
limit. Following this procedure, not more than 10\% of the galaxies
could be missed in the lowest mass end of the mass function.

At $z>2$, galaxies could be too faint to be detected in optical and we
need to rely on optical upper-limits and NIR fluxes to estimate the
photo-z. Photometric redshifts with a precision of 5\% are routinely
computed with four bands (e.g. Ilbert et al. 2006). Therefore, we
checked that most of the galaxies are detected in four bands at least,
to allow a robust photo-z estimate.  The fraction of galaxies detected
in a minimum of four bands at 2 and 5 $\sigma$ are given in Figure
\ref{completeness}. This fraction is always above 95\% at $z<2$ even
if we consider a 5$\sigma$ detection limit. At $z>2$, the fraction
stays higher than 80\% (90\%) for a 5$\sigma$ (2$\sigma$) detection
limit, except at $2.5<z<3$ for the quiescent (still 91\% have their
magnitudes measured in a minimum of 4 bands, with a detection limit
better than 2$\sigma$). Moreover, as indicated in Figure
  \ref{completeness}, more than $75\%$ of sources show a 5$\sigma$
  detection at 3.6$\mu m$ even at the highest redshift bin.

   \begin{figure}
   \centering
   \includegraphics[width=9.3cm]{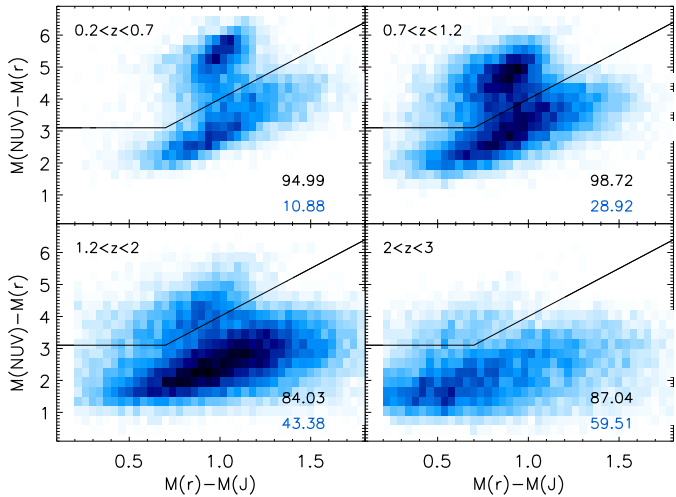}
   \caption{Two-colour selection of the quiescent population. The
     galaxies above the black line in the top left are selected
     as quiescent. The fraction (in \%) of $log(sSFR)<-11$ galaxies
     selected by the two-colour selection box is the top number in
     each panel. The fraction of galaxies with $log(sSFR)>-11$ within
     the two-colour box is the bottom number.}
              \label{color}%
    \end{figure}

\subsection{Galaxy classification}\label{class}

In order to divide the sample into quiescent and star-forming
galaxies, we use a slightly modified version of the two-colour
selection technique proposed by Williams et al. (2009). Following
Ilbert et al. (2010), we use the rest-frame two-colour selection ${\rm
  NUV} -r^+$ versus $r^+-J$ instead of $U-V$ versus $V-J$. In fact,
${\rm NUV} -r^+$ is a better indicator of the current versus past star
formation activity (e.g. Martin et al. 2007, Arnouts et
al. 2007)\footnote{${\rm NUV}$ corresponds to the GALEX filter
  centered at $0.23 \mu m$}. Moreover, the dynamical range covered by
the ${\rm NUV} -r^+$ rest-frame colour is larger than the one covered
by $U-V$, making the ${\rm NUV} -r^+$ rest-frame colour less sensitive
to uncertainties linked to observations. Finally, the ${\rm NUV}$
rest-frame is still sampled by optical data at $z>2$ which is no
longer true for the rest frame $U$ band.

Figure \ref{color} shows the two-colour criterion used to select the
quiescent population. Galaxies with $M_{NUV}-M_r > 3(M_r-M_J)+1$ and
$M_{NUV}-M_r > 3.1$ are considered as quiescent. The advantage of this
classification is that it avoids a mix between dusty star-forming
galaxies and quiescent galaxies: extinction moves star-forming
galaxies along a diagonal axes from the bottom left to the top right
of Figure \ref{color}. We derive the absolute magnitudes with the
method of Ilbert et al. (2005) which minimizes the k-correction
dependency\footnote{When the absolute magnitude is computed at
  $\lambda_{rest-frame}$, we base its estimate on the apparent
  magnitude measured at $\lambda_{rest-frame}(1+z_{gal})$}.

The specific star formation rate (sSFR, estimated in $yr^{-1}$
  throughout the paper) is the ratio between the instantaneous SFR
and the stellar mass obtained from the best-fit template.  In Ilbert
et al. (2010) and Dom{\'{\i}}nguez S{\'a}nchez et al. (2011), we
considered that a galaxy was quiescent when $log(sSFR)<-11$. The
fraction of galaxies with $log(sSFR)<-11$ in the selection box is
given in each panel of Figure \ref{color}.  Over the full redshift
range, $\sim$90\% of the galaxies with $log(sSFR)<-11$ are in the
selection box. We also indicate the fraction of galaxies with
$log(sSFR)>-11$ in the selection box. The fraction of galaxies with
$log(sSFR)>-11$ increases from 20\% at $z<1$ to 60\% at
$2<z<2.5$. Therefore, both classifications provide similar results at
$z<1$. But the classification based on the sSFR is more conservative
at high redshift.

\begin{table*}[htb!]
\begin{center}
\begin{tabular}{l c c c c c c c c c} \hline

             &          &         &      $log({\cal M}_{complete})$                &   $log({\cal M}^*)$ &    $\Phi^*_1$                 &    $\alpha_1$                   &    $\Phi^*_2$                  &    $\alpha_2$      & $log(\rho^*)$  \\ 
type         &   z-bin  &  Number & (${\cal M}_{\sun}$)  &   (${\cal M}_{\sun}$) &  ($10^{-3} Mpc^{-3}$)          &                                 &   ($10^{-3} Mpc^{-3}$)           &                    & (${\cal M}_{\sun} Mpc^{-3}$) \vspace{0.2cm}\\  \hline
\hline \\

full     & 0.2-0.5 &   26650 &       7.93  &   10.88$^{{\rm +  0.10}}_{{\rm  -0.10}}$ &    1.68$^{{\rm +  0.61}}_{{\rm  -0.61}}$ &   -0.69$^{{\rm +  0.40}}_{{\rm  -0.36}}$ &    0.77$^{{\rm +  0.40}}_{{\rm  -0.53}}$ &   -1.42$^{{\rm +  0.07}}_{{\rm  -0.14}}$  &  $8.308^{\rm +0.080}_{\rm -0.095}$ \\
sample   & 0.5-0.8 &   29418 &       8.70  &   11.03$^{{\rm +  0.08}}_{{\rm  -0.10}}$ &    1.22$^{{\rm +  0.31}}_{{\rm  -0.39}}$ &   -1.00$^{{\rm +  0.31}}_{{\rm  -0.31}}$ &    0.16$^{{\rm +  0.32}}_{{\rm  -0.32}}$ &   -1.64$^{{\rm +  0.20}}_{{\rm  -0.50}}$  &  $8.226^{\rm +0.065}_{\rm -0.073}$ \\
         & 0.8-1.1 &   27590 &       9.13  &   10.87$^{{\rm +  0.06}}_{{\rm  -0.06}}$ &    2.03$^{{\rm +  0.27}}_{{\rm  -0.32}}$ &   -0.52$^{{\rm +  0.35}}_{{\rm  -0.27}}$ &    0.29$^{{\rm +  0.30}}_{{\rm  -0.30}}$ &   -1.62$^{{\rm +  0.19}}_{{\rm  -0.32}}$  &  $8.251^{\rm +0.069}_{\rm -0.073}$ \\
         & 1.1-1.5 &   29383 &       9.42  &   10.71$^{{\rm +  0.08}}_{{\rm  -0.08}}$ &    1.35$^{{\rm +  0.34}}_{{\rm  -0.35}}$ &   -0.08$^{{\rm +  0.55}}_{{\rm  -0.52}}$ &    0.67$^{{\rm +  0.41}}_{{\rm  -0.44}}$ &   -1.46$^{{\rm +  0.16}}_{{\rm  -0.29}}$  &  $8.086^{\rm +0.069}_{\rm -0.071}$ \\
         & 1.5-2.0 &   20529 &       9.67  &   10.74$^{{\rm +  0.07}}_{{\rm  -0.06}}$ &    0.88$^{{\rm +  0.10}}_{{\rm  -0.12}}$ &   -0.24$^{{\rm +  0.27}}_{{\rm  -0.28}}$ &    0.33$^{{\rm +  0.06}}_{{\rm  -0.07}}$ &   -1.6  &  $7.909^{\rm +0.090}_{\rm -0.072}$ \\
         & 2.0-2.5 &    7162 &      10.04  &   10.74$^{{\rm +  0.07}}_{{\rm  -0.07}}$ &    0.62$^{{\rm +  0.07}}_{{\rm  -0.07}}$ &   -0.22$^{{\rm +  0.29}}_{{\rm  -0.29}}$ &    0.15$^{{\rm +  0.04}}_{{\rm  -0.04}}$ &   -1.6  &  $7.682^{\rm +0.110}_{\rm -0.081}$ \\
         & 2.5-3.0 &    3143 &      10.24  &   10.76$^{{\rm +  0.16}}_{{\rm  -0.15}}$ &    0.26$^{{\rm +  0.05}}_{{\rm  -0.08}}$ &   -0.15$^{{\rm +  0.86}}_{{\rm  -0.68}}$ &    0.14$^{{\rm +  0.11}}_{{\rm  -0.06}}$ &   -1.6  &  $7.489^{\rm +0.230}_{\rm -0.123}$ \\
         & 3.0-4.0 &    1817 &      10.27  &   10.74$^{{\rm +  0.44}}_{{\rm  -0.20}}$ &    0.03$^{{\rm +  0.02}}_{{\rm  -0.02}}$ &    0.95$^{{\rm +  1.05}}_{{\rm  -1.21}}$ &    0.09$^{{\rm +  0.07}}_{{\rm  -0.07}}$ &   -1.6  &  $7.120^{\rm +0.234}_{\rm -0.168}$ 
\vspace{0.1cm}\\
\hline \\

quiescent        &  0.2-0.5 &   3878  &   8.24  &  10.91$^{{\rm +  0.07}}_{{\rm  -0.08}}$ &    1.27$^{{\rm +  0.21}}_{{\rm  -0.19}}$ &   -0.68$^{{\rm +  0.22}}_{{\rm  -0.13}}$ &    0.03$^{{\rm +  0.07}}_{{\rm  -0.07}}$ &   -1.52$^{{\rm +  0.27}}_{{\rm  -0.44}}$ &  7.986$^{\rm +0.087}_{\rm -0.102 }$  \\
galaxies         &  0.5-0.8 &   3910  &   8.96  &  10.93$^{{\rm +  0.04}}_{{\rm  -0.04}}$ &    1.11$^{{\rm +  0.10}}_{{\rm  -0.09}}$ &   -0.46$^{{\rm +  0.05}}_{{\rm  -0.05}}$ &                &      &                                                                7.920$^{\rm +0.054}_{\rm -0.058 }$  \\
                 &  0.8-1.1 &   5591  &   9.37  &  10.81$^{{\rm +  0.03}}_{{\rm  -0.03}}$ &    1.57$^{{\rm +  0.09}}_{{\rm  -0.09}}$ &   -0.11$^{{\rm +  0.05}}_{{\rm  -0.05}}$ &                &      &                                                                7.985$^{\rm +0.044}_{\rm -0.049 }$  \\
                 &  1.1-1.5 &   3769  &   9.60  &  10.72$^{{\rm +  0.03}}_{{\rm  -0.03}}$ &    0.70$^{{\rm +  0.03}}_{{\rm  -0.03}}$ &    0.04$^{{\rm +  0.06}}_{{\rm  -0.06}}$ &                &      &                                                                7.576$^{\rm +0.041}_{\rm -0.046 }$  \\
                 &  1.5-2.0 &   1612  &   9.87  &  10.73$^{{\rm +  0.03}}_{{\rm  -0.04}}$ &    0.22$^{{\rm +  0.01}}_{{\rm  -0.01}}$ &    0.10$^{{\rm +  0.09}}_{{\rm  -0.09}}$ &                &      &                                                                7.093$^{\rm +0.049}_{\rm -0.053 }$  \\
                 &  2.0-2.5 &    800  &  10.11  &  10.59$^{{\rm +  0.06}}_{{\rm  -0.06}}$ &    0.10$^{{\rm +  0.01}}_{{\rm  -0.01}}$ &    0.88$^{{\rm +  0.23}}_{{\rm  -0.21}}$ &                &      &                                                                6.834$^{\rm +0.076}_{\rm -0.084 }$  \\
                 &  2.5-3.0 &    240  &  10.39  &  10.27$^{{\rm +  0.10}}_{{\rm  -0.08}}$ &    0.003$^{{\rm +  0.006}}_{{\rm  -0.002}}$ & 3.26$^{{\rm +  0.93}}_{{\rm  -0.93}}$ &                &      &                                                                6.340$^{\rm +0.079}_{\rm -0.121 }$  
\vspace{0.1cm}\\
\hline \\

 star-forming  & 0.2-0.5  &    23124 &   7.86   &   10.60$^{{\rm +  0.16}}_{{\rm  -0.11}}$ &    1.16$^{{\rm +  0.31}}_{{\rm  -0.41}}$ &    0.17$^{{\rm +  0.57}}_{{\rm  -0.65}}$ &    1.08$^{{\rm +  0.29}}_{{\rm  -0.31}}$ &   -1.40$^{{\rm +  0.04}}_{{\rm  -0.04}}$ &  $8.051^{\rm +0.091}_{\rm -0.101}$ \\
  galaxies     & 0.5-0.8  &    26830 &   8.64   &   10.62$^{{\rm +  0.17}}_{{\rm  -0.10}}$ &    0.77$^{{\rm +  0.22}}_{{\rm  -0.30}}$ &    0.03$^{{\rm +  0.58}}_{{\rm  -0.79}}$ &    0.84$^{{\rm +  0.28}}_{{\rm  -0.35}}$ &   -1.43$^{{\rm +  0.06}}_{{\rm  -0.09}}$ &  $7.933^{\rm +0.069}_{\rm -0.073}$ \\
               & 0.8-1.1  &    24184 &   9.04   &   10.80$^{{\rm +  0.11}}_{{\rm  -0.12}}$ &    0.50$^{{\rm +  0.33}}_{{\rm  -0.31}}$ &   -0.67$^{{\rm +  0.75}}_{{\rm  -0.67}}$ &    0.48$^{{\rm +  0.41}}_{{\rm  -0.41}}$ &   -1.51$^{{\rm +  0.11}}_{{\rm  -0.67}}$ &  $7.908^{\rm +0.065}_{\rm -0.067}$ \\
               & 1.1-1.5  &    29934 &   9.29   &   10.67$^{{\rm +  0.11}}_{{\rm  -0.09}}$ &    0.53$^{{\rm +  0.23}}_{{\rm  -0.18}}$ &    0.11$^{{\rm +  0.61}}_{{\rm  -0.78}}$ &    0.87$^{{\rm +  0.29}}_{{\rm  -0.40}}$ &   -1.37$^{{\rm +  0.08}}_{{\rm  -0.15}}$ &  $7.916^{\rm +0.059}_{\rm -0.061}$ \\
               & 1.5-2.0  &    19570 &   9.65   &   10.66$^{{\rm +  0.07}}_{{\rm  -0.06}}$ &    0.75$^{{\rm +  0.08}}_{{\rm  -0.10}}$ &   -0.08$^{{\rm +  0.28}}_{{\rm  -0.31}}$ &    0.39$^{{\rm +  0.07}}_{{\rm  -0.07}}$ &   -1.6 &  $7.841^{\rm +0.097}_{\rm -0.071}$ \\
               & 2.0-2.5  &     6597 &   10.01  &   10.73$^{{\rm +  0.08}}_{{\rm  -0.08}}$ &    0.50$^{{\rm +  0.07}}_{{\rm  -0.07}}$ &   -0.33$^{{\rm +  0.33}}_{{\rm  -0.33}}$ &    0.15$^{{\rm +  0.05}}_{{\rm  -0.05}}$ &   -1.6 &  $7.614^{\rm +0.123}_{\rm -0.084}$ \\
               & 2.5-3.0  &     3035 &   10.20  &   10.90$^{{\rm +  0.20}}_{{\rm  -0.21}}$ &    0.15$^{{\rm +  0.08}}_{{\rm  -0.08}}$ &   -0.62$^{{\rm +  1.04}}_{{\rm  -0.84}}$ &    0.11$^{{\rm +  0.07}}_{{\rm  -0.07}}$ &   -1.6 &  $7.453^{\rm +0.193}_{\rm -0.128}$ \\
               & 3.0-4.0  &     1829 &   10.26  &   10.74$^{{\rm +  0.29}}_{{\rm  -0.17}}$ &    0.02$^{{\rm +  0.01}}_{{\rm  -0.01}}$ &    1.31$^{{\rm +  0.87}}_{{\rm  -0.87}}$ &    0.10$^{{\rm +  0.06}}_{{\rm  -0.04}}$ &   -1.6 &  $7.105^{\rm +0.245}_{\rm -0.170}$ \\

\end{tabular}
\caption{Best-fit parameters for the full sample, for the quiescent
  galaxies classified with the two-colour criteria and for the
  star-forming galaxies. We adopt a double Schechter function (see
  Eq.2) convolved with stellar mass uncertainties (see Appendix A) to
  fit the {\it Vmax} data points. A simple Schechter function is
  considered for the quiescent sample at $z>0.5$. The slope $\alpha_2$ is set at
  -1.6 for star-forming galaxies and the full sample at $z>1.5$. We allow the
  slope to vary between $-1.9<\alpha_2<-1.4$ to compute the 
  mass density uncertainties. \label{schechter}}
\end{center}
\end{table*}

\begin{figure}
\centering \includegraphics[width=9cm]{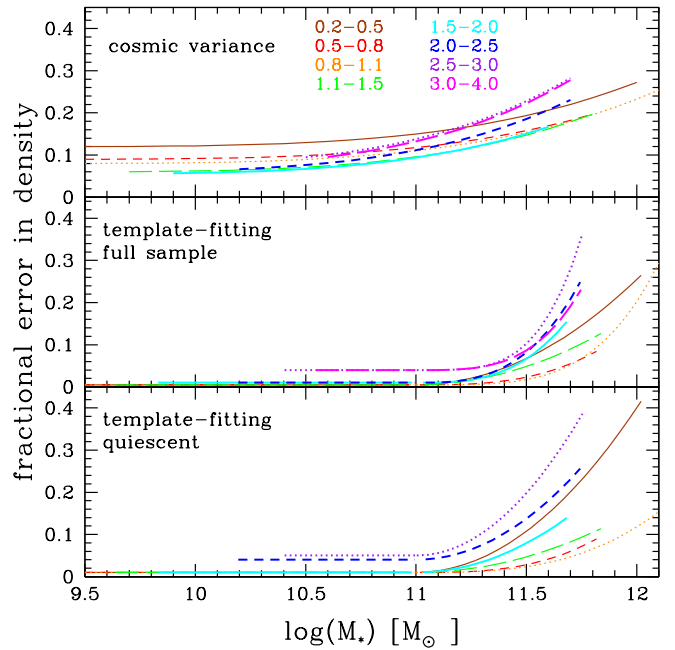}
\caption{Fractional error in density as a function of the stellar
  mass and redshift. The top panel shows the errors due to the
  cosmic variance. The middle and bottom panels are the errors
  associated to the template fitting procedure (photo-z and stellar mass) for
  the full sample and the quiescent population,
  respectively. Results are shown only in the mass range covered by our
  dataset.}
           \label{error} 
\end{figure}

\section{Method used to estimate the Galaxy Stellar Mass Function}\label{estimate}

We use the tool ALF (Ilbert et al. 2005) to derive the galaxy stellar
mass functions. This tool includes three non-parametric estimators
(Vmax, SWML, C$^+$) detailed in Appendix A.2 of Ilbert et
al. (2005). We limit our analysis to above the stellar mass limit given
in Table 2 (column 4), as defined in \S\ref{stellarmasses}.  We
verified that the three non-parametric estimators are in good agreement
over the considered mass ranges. As shown in Ilbert et al. (2004), the
estimators behave differently if a galaxy population is missing and
diverge below a given mass limit when a significant population is no
longer present. The good agreement between the three estimators
confirm that our stellar mass limits are well established.

\subsection{Associated uncertainties}

Our error budget includes the Poissonian errors, the photo-z redshift
uncertainties, and the uncertainties in the mass estimations.

We estimate the cosmic variance with the public tool \texttt{getcv}
provided by Moster et al. (2011). The square root of the cosmic
variance for galaxies is computed as the product of the galaxy bias
and the dark matter root cosmic variance. The dark matter cosmic
variance is computed for a field of $1.5~\mathrm{deg}^2$. Moster et
al. (2011) provide an estimate of the bias as a function of redshift
and stellar mass. The fractional error in density is shown in the top
panel of Figure \ref{error}. This uncertainty increases with the mass
since the massive galaxies are more biased than lower mass galaxies.
The error associated to the cosmic variance is always larger than 5\%
at low mass and reaches 30\% at high mass.

In order to take into account the uncertainties generated by the
template-fitting procedure (photo-z and stellar mass), we created a
set of 30 mock catalogues by perturbing each flux point according to
its formal error bar. We scatter the photometric redshifts according
to their 1$\sigma$ error provided by \texttt{Le\_Phare}. For each
realisation, we recompute the stellar masses, and perform again the
galaxy classification and recompute the MFs. We compute the 1$\sigma$
dispersion in density at a given mass and redshift. The evolution of
this dispersion is shown in Figure \ref{error}. These uncertainties
are below 2\% at low masses and $z<2$. In this regime, the cosmic
variance dominates the error budget. At high masses ${\cal
  M}>10^{11}{\cal M}_{\sun}$, the errors associated with the template
fitting procedure reach 30-40\% and become as important as the cosmic
variance. 

In order to obtain the total errors, we add in quadrature
the errors due to the galaxy cosmic variance ($err_{cosmic\_var}$),
the ones linked to the template-fitting procedure ($err_{fit}$) and
the Poissonian errors ($err_{poisson}$):
\begin{equation}
err_{tot}=\sqrt{err_{poisson}^2+err_{cosmic\_var}^2+err_{fit}^2}.
\end{equation}
Since $err_{cosmic\_var}$ and $err_{fit}$ are derived as a function of
the mass and the redshift, it is straightforward to associate the
total errors to the non-parametric MF data points.

\subsection{Fit of the stellar mass function and Eddington bias}

We fit a parametric form over the {\it Vmax} non-parametric data. The
choice of the non-parametric estimator has no impact on our results
since we work in a mass range where the three non-parametric
estimators are consistent.  Following Pozzetti et al. (2010), we use a
double Schechter function, defined as:
\begin{equation}
 \phi({\cal M})d{\cal
    M}=e^{-\frac{{\cal M}}{{\cal M}^*}}\left[\phi^*_1\left(\frac{{\cal M}}{{\cal
        M}^*}\right)^{\alpha_1}+\phi^*_2\left(\frac{{\cal M}}{{\cal
        M}^*}\right)^{\alpha_2}\right]\frac{d{\cal M}}{{\cal M}^*}
\end{equation}

with $\cal M^*$ the characteristic stellar mass, $\alpha_1$ and
$\alpha_2$ are the slopes which need to satisfy the following criteria
$\alpha_2<\alpha_1$, and $\phi^*_1$ and $\phi^*_2$ correspond to the
normalization. For the global and star-forming populations, we
arbitrary set the slope at $\alpha_2=-1.6$ at $z>1.5$ since this
parameter is no longer constrained. For the quiescent population, we
use a simple Schechter function at $z>0.5$ since we do not detect any
upturn at the low-mass end given our survey limit. At $0.2<z<0.5$, we
still need to use a fit with a double Schechter function.

Below the stellar mass limit ${\cal M}_{complete}$ given in Table 2
(see \S\ref{stellarmasses}), the {\it Vmax} estimator underestimates
the true MF (Ilbert et al. 2004). Therefore, we use the {\it Vmax}
measurements below ${\cal M}_{complete}$ as a lower bound in the
fitting procedure.

A crucial step in our fitting procedure is to account for the
uncertainties in the stellar mass. These uncertainties could bias our
estimate of the high-mass end (Caputi et al. 2011). Since the galaxy
density exponentially decreases towards massive galaxies, errors in
the stellar mass scatters more galaxies into the massive end than the
reverse (Eddington 1913). Our procedure to avoid this bias is detailed
in Appendix A. First, we find that the stellar mass uncertainties are
well characterised by the product of a Lorentzian distribution $
L(x)=\frac{\tau}{2\pi}\frac{1}{(\frac{\tau}{2})^2+ x^2}$ with
$\tau=0.04(1+z)$ and a Gaussian distribution $G$ with
$\sigma=0.5$. Then, we convolve the double Schechter function $\phi$
by the stellar mass uncertainties: $ \phi_{convolved} = \phi * (L
\times G)$. Finally, we fit $ \phi_{convolved}$ to the {\it Vmax}
non-parametric data.  Therefore, the best-fit parameters that we
provide in Table 2 are deconvolved by the expected stellar mass
uncertainties and do not suffer from Eddington bias.

\section{Results: Evolution of the Galaxy Stellar Mass Function and Stellar Mass Density}\label{MF}

\begin{figure}
   \centering
   \includegraphics[width=9cm]{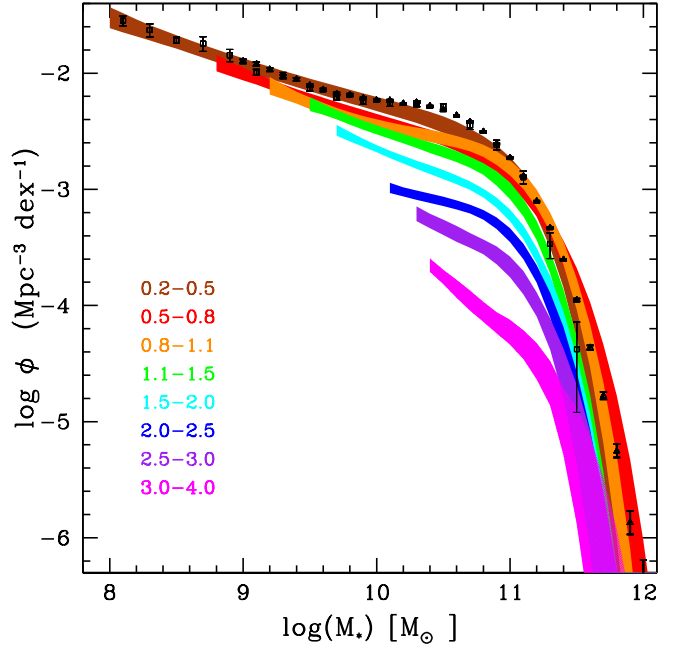}
   \caption{Galaxy stellar mass function up to $z=4$ for the full
     sample. Each colour corresponds to different redshift bins of
     variable step size. Fits are shown in the mass range covered by
     our dataset. The filled areas correspond to the 68\% confidence
     level regions, after accounting for Poissonian errors, the cosmic
     variance and the uncertainties created during the template
     fitting procedure. The open triangles and squares correspond to
     the local estimates by Moustakas et al. (2013) and Baldry et
     al. (2012), respectively. }
              \label{MFone}%
    \end{figure}

\begin{figure}
   \centering
   \includegraphics[width=9cm]{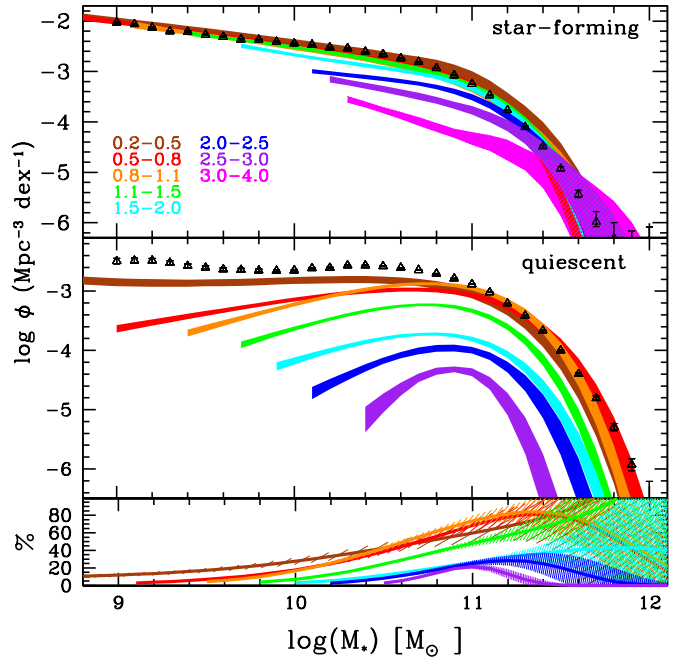}
   \caption{Galaxy stellar mass function up to $z=4$ for the
     star-forming population (top panel) and for the quiescent
     population (middle panel). Symbols are the same as Figure
     \ref{MFone}. The bottom panel shows the percentage of quiescent
     galaxies as a function of stellar mass in the same redshift
     bins.}
              \label{MFtype}%
    \end{figure}

The galaxy stellar mass functions are computed with a sample of
220,000 galaxies selected at $K_{\rm s}<24$.  We keep only the sources
in areas with good image quality, representing an area of 1.52
deg$^2$. We remove the stars and X-ray detected AGNs (Brusa et
al. 2007). Figure \ref{MFone} and Figure \ref{MFtype} show the galaxy
stellar mass functions for the full sample, the quiescent and the
star-forming populations. The best fit parameters are given in Table
2. In this Section, we describe our results out to $z=4$.

   \begin{figure*}
   \centering
   \includegraphics[width=17cm]{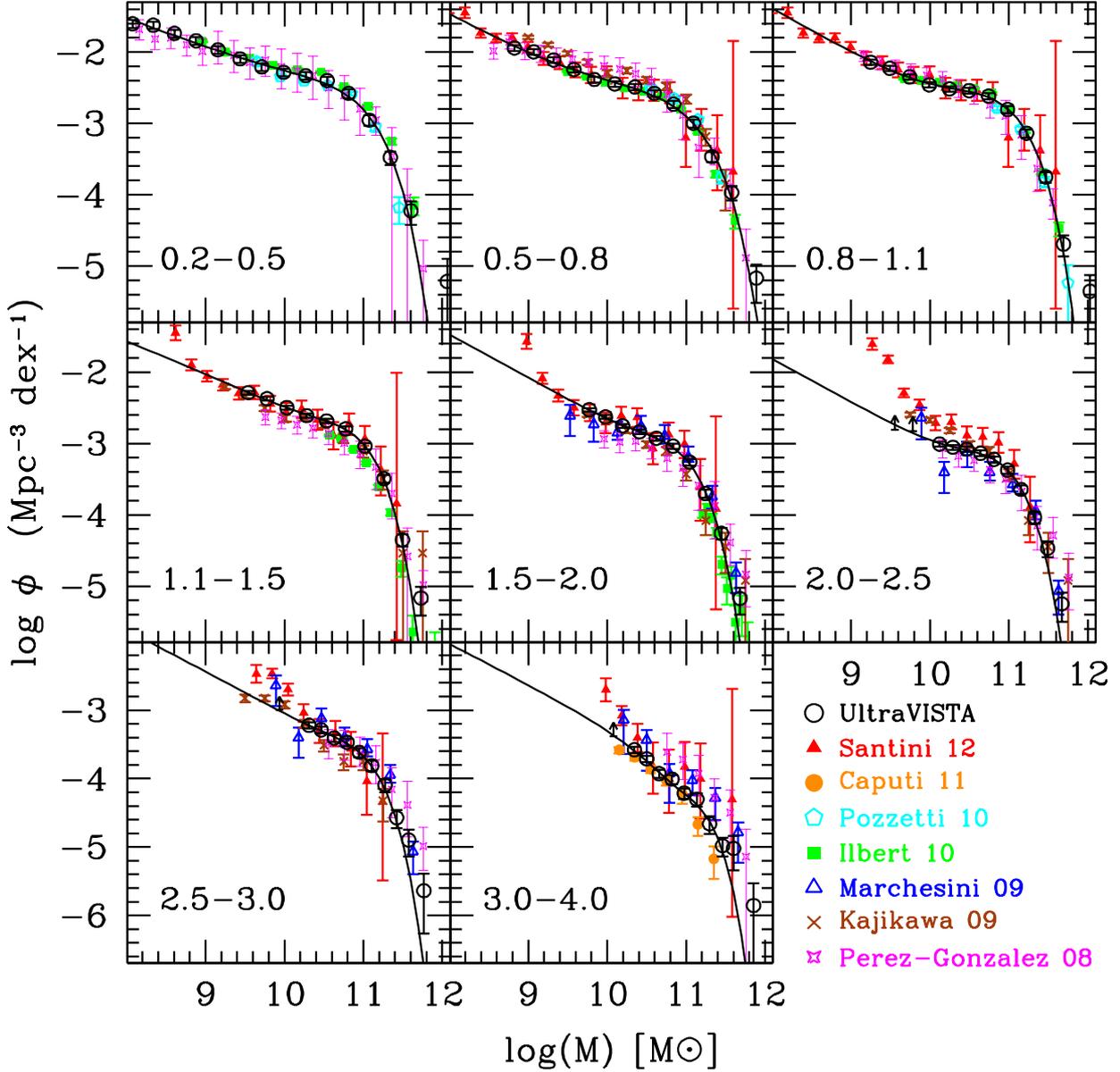}
   \caption{UltraVISTA global stellar mass functions (open black
       circles and solid lines) compared with several measurements
       from the literature published since 2008 (labeled in the bottom
       right). Each panel corresponds to a redshift bin. The
       literature MF measurements are converted to the same cosmology
       and IMF as used in this study
       ($H_{\rm0}~=~70$~km~s$^{-1}$~Mpc$^{-1}$ and Chabrier 2003
       IMF).}
              \label{MFlitt}%
    \end{figure*}

   \begin{figure}
   \centering
   \includegraphics[width=9cm]{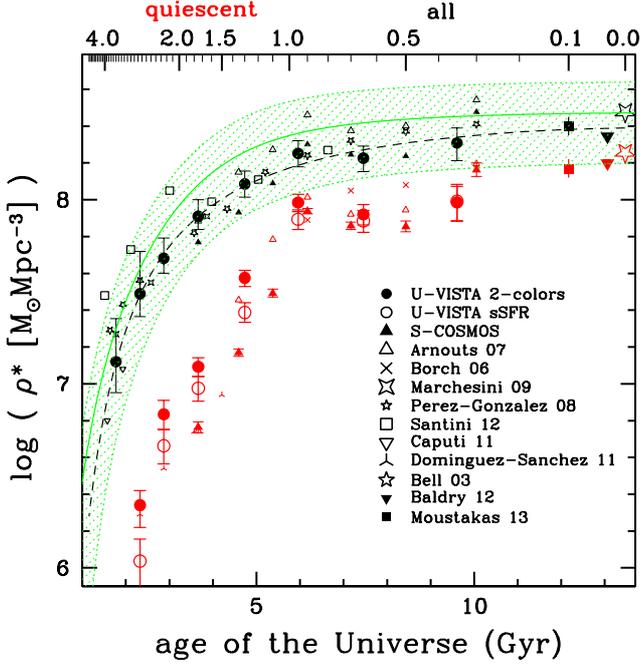}
   \caption{Stellar mass density as a function of cosmic time
     (redshift is given in the top label). Black and red points
     correspond to the full and quiescent populations,
     respectively. The circles correspond to our new results using
     UltraVISTA. Solid and open red circles correspond to the
     two-colour and sSFR selected quiescent galaxies,
     respectively. The green shaded area corresponds to the cosmic SFR
     compiled by Behroozi et al. (2013) and integrated over cosmic
     time as described in section \ref{SFR}. The dashed line
       corresponds to the best fit over the mass density data.}
              \label{LD}%
    \end{figure}

   \begin{figure}
   \centering
   \includegraphics[width=9cm]{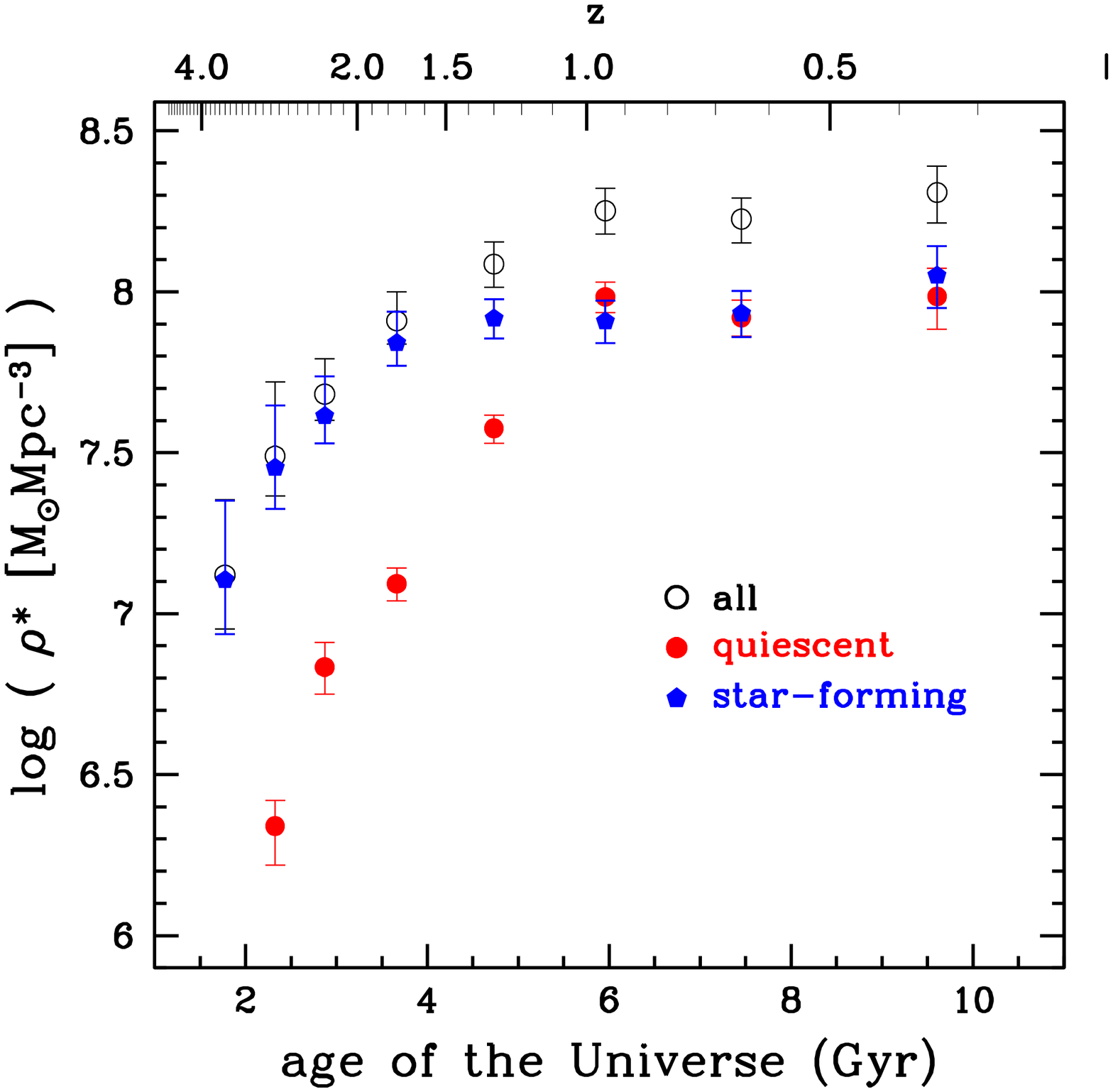}
   \caption{Stellar mass density as a function of cosmic time. Black
     open circles, red solid circles and blue pentagons correspond to
     the full, quiescent and star-forming galaxies, respectively. }
              \label{LD2}%
    \end{figure}

\subsection{Evolution of the full sample}\label{full}

Figure \ref{MFone} shows the evolution of the MFs for the full
sample. A first option is to consider a pure evolution in stellar
mass. In this case, we assume that only star formation drives the MF
evolution (no galaxy can be created or destroyed). We find that the
evolution is strongly mass-dependent, with the low-mass end evolving
more rapidly than the high-mass end. For instance, the stellar mass of
a $10^{9.8} {\cal M}_{\sun}$ galaxy increases by 0.9 dex between
$1.5<z<2$ and $0.2<z<0.5$, while the stellar mass of a $10^{11.6}
{\cal M}_{\sun}$ galaxy increases by only 0.2 dex in the same time
interval. Therefore, we conclude that the evolution is strongly
mass-dependent, in agreement Marchesini et al. (2009). A second option
is to consider a pure density evolution. A constant increase in
density by 0.3-0.4 dex, independent of the mass, is sufficient to
match the $1.5<z<2$ and the $0.2<z<0.5$ MFs. However such a pure
density evolution scenario is not applicable to the full sample: it
would mean that new galaxies which were not present in a given
redshift appear in the next redshift bin. Major mergers are not an
option for a pure increase in density with cosmic time: for a
$\alpha=-1.4$ MF slope, the density of low mass galaxies would
decrease by 0.16 dex if we assume that all galaxies encounter a major
merger since $z=2$\footnote{The MF would be shifted in density by -0.3
  dex (half as many galaxies) and the masses would increase by 0.3
  dex}. 

In Figure \ref{MFlitt}, we compare our results with several MF
estimates published since 2008. We find an excellent agreement with
the various MFs from the literature. Still, the differences in
normalisation are as large as 0.2 dex in certain bins
(e.g. $0.5<z<0.8$ with Kajikawa et al. (2009) and P\'erez-Gonz\'alez
et al. (2008); at $2<z<2.5$ with Santini et al. (2012) which could be
explained by known groups at $z\sim 2.2-2.3$). We also find that the
extrapolation of our MF slope is flatter than data from Santini et
al. (2012), but our sample does not reach a similar depth as this
study.

We derive the stellar mass density by integrating the best-fit double
Schechter functions over the mass range $10^{8}$ to $10^{13} {\cal
  M}_{\sun}$. Since our mass limits are above $10^{10} {\cal
    M}_{\sun}$ at $z>2$ (see Table 2), our mass density estimates rely
  on the slope extrapolation for low mass galaxies. Since the slope is
  arbitrary set at $\alpha_2=-1.6$ at $z>1.5$, we allow $\alpha_2$ to
  vary in the range $-1.9<\alpha_2<-1.4$ to compute the mass density
  uncertainties. The results are shown in Figure \ref{LD} (black
circles). We fit the stellar mass density by the parametric form
  $\rho_*(z) = a \times e^{ - bz^c}$. Our best fit function is shown
  with a dashed line in Figure \ref{LD} for the best fitting
  parameters $a=2.46^{+0.35}_{-0.29} \times 10^8$,
  $b=0.50^{+0.19}_{-0.16}$ and $c=1.41^{+0.40}_{-0.31}$. We find that
the global stellar mass density increases by 1.1 dex between $3<z<4$
and $0.8<z<1.1$ (a factor 13 in 4.2 Gyr). The evolution is much
smaller at $z<1$ with an increase of 0.2 dex between $0.8<z<1.1$ and
the local estimate (a factor 1.6 in 6 Gyr). Therefore, the stellar
mass is assembled twice as fast at $1<z<4$ (14\% of the local stellar
mass density per Gyr) than at $z<1$ (6\% of the local stellar mass
density per Gyr).

Figure \ref{LD} shows an excellent agreement between the various
estimates from the literature, in particular at high redshift with the
previous results from Caputi et al. (2011), P\'erez-Gonz\'alez et
al. (2008) and Marchesini et al. (2009). We derive stellar mass
densities which are lower by 0.2-0.3 dex at $z>2$ in comparison with
Santini et al. (2012). They find a slope much steeper than ours
($\alpha < -1.8$) which explains this difference. Their steep slopes
are explained by a steep upturn in their faintest bins in stellar mass
(see the red triangles in Figure \ref{MFlitt}). Our MF and mass
density are in good agreement if we limit the comparison to the mass
range common to both studies. Santini et al. use a HAWK-I K$_{\rm s}$
selected sample which is 1.5 magnitude deeper than our sample and we
are not able to reach the stellar mass limit at which this upturn
occurs.

\subsection{Evolution of the quiescent population}\label{ell}

Middle panel of Figure \ref{MFtype} shows the MF evolution of
quiescent galaxies. The MF evolution of the quiescent population is
clearly mass dependent at $z<1$. In this redshift range, we do not
find any significant evolution of the high-mass end while we observe a
clear flattening of the faint-end slope.  Between $0.8<z<1.1$ and
$0.2<z<0.5$, the density of galaxies more massive than $10^{11.2}
{\cal M}_{\sun}$ does not increase, while galaxies are continuously
``quenched'' at the low-mass end. For instance, the density of
$10^{9.5} {\cal M}_{\sun}$ galaxies increases by a factor of $>5$
between $0.8<z<1.1$ and $0.2<z<0.5$.

At $1<z<3$, we find a rapid increase in density of all quiescent
galaxies. In contrast with the result at $z<1$, the evolution is no
longer mass dependent. The most massive galaxies evolve as fast as
intermediate mass galaxies. The density of $10^{11} {\cal M}_{\sun}$
galaxies increases by 1.4 dex (factor 25) between $2.5<z<3$ and
$0.8<z<1.1$. The normalisation parameter $\Phi^*_1$ increases
continuously between $z\sim 3$ and $z\sim 1$. We consider a pure
density evolution for the quiescent MF rather than a pure mass
evolution. Indeed, new quiescent galaxies are created along cosmic
time by quenching of star-forming galaxies. Moreover, an isolated
quiescent galaxy grows by less than 0.03 dex in 6 Gyr (since
$log(sSFR)<-11$) due to its own star-formation.

Figure \ref{LD} and Figure \ref{LD2} show the evolution of the stellar
mass density for the quiescent population. The stellar mass density
increases by 1.6 dex between $2.5<z<3$ and $0.8<z<1.1$. Again, we find
a change of regime around $z\sim 1$ with the mass assembly slowing
down at $z<1$. We find that the stellar mass assembly is faster at
$1<z<3$ for the quiescent population than for the global population
which evolves by 0.8 dex in the same redshift range. For the massive
quiescent galaxies (${\cal M} > 10^{11}{\cal M}_{\sun}$), Brammer et
al. (2011) find an evolution of 0.5 dex between $z=2.1$ and $z=1$. But
their mass density drops quickly at $z>2.1$. Their evolution reaches 1
dex if we consider the redshift range $z=2.3$ and $z=1$, which is
really close to our value of 1.1 dex. We find a slightly slower
evolution compared to Ilbert et al. (2010). However, our conclusion
depends on the method used to classify the quiescent population. If we
use a classification based on the sSFR, as in Ilbert et al. (2010),
the stellar mass densities are consistent between both studies. The
sSFR classification is more restrictive than the two-colour selection
(see \S\ref{class}).

\subsection{Evolution of the star-forming population}

The top panel of Figure \ref{MFtype} shows the stellar mass function
of the star-forming galaxies. We can divide this evolution in two
regimes: above and below $10^{10.7-10.9} {\cal M}_{\sun}$. If we
consider only the low mass regime, the faint-end slope remains steep
over the full redshift range. We do not detect any significant trend
in the slope evolution over the stellar mass range covered by our
dataset. We observe a strong evolution of low-mass galaxies,
especially at $z>2$. For the MF at $2<z<2.5$ to match that at
$1.5<z<2$ requires the stellar mass of a $10^{10.3} {\cal M}_{\sun}$
galaxy to increase by around 0.4 dex. We discuss in \S\ref{ssfr} how
such evolution could be interpreted in term of sSFR.  If we consider
now the evolution of the high-mass end, we do not detect an evolution
of the density of the most massive galaxies $10^{11.6-11.8} {\cal
  M}_{\sun}$. Since these star-forming galaxies are forming new
stellar populations, these massive star-forming galaxies are
necessarily quenched along cosmic time, as we will discuss in
\S\ref{quenching}.

Figure \ref{LD2} shows the evolution of the stellar mass density for
star-forming galaxies. As for the quiescent population, we also
observe two regimes with a faster evolution at $1<z<4$ than at
$z<1$. We observe an evolution of 0.5 dex between $2.5<z<3$ and
$0.8<z<1.1$, while the quiescent galaxies evolve by 1.6 dex in the
same period. Therefore, quiescent galaxies are building faster at this
epoch. We also observe that the stellar mass density of quiescent and
star-forming galaxies are comparable at $z<1$, while star-forming
galaxies dominate the stellar mass budget at higher redshift.

\section{Discussion}\label{discussion}

\begin{figure}
\centering
\includegraphics[width=9cm]{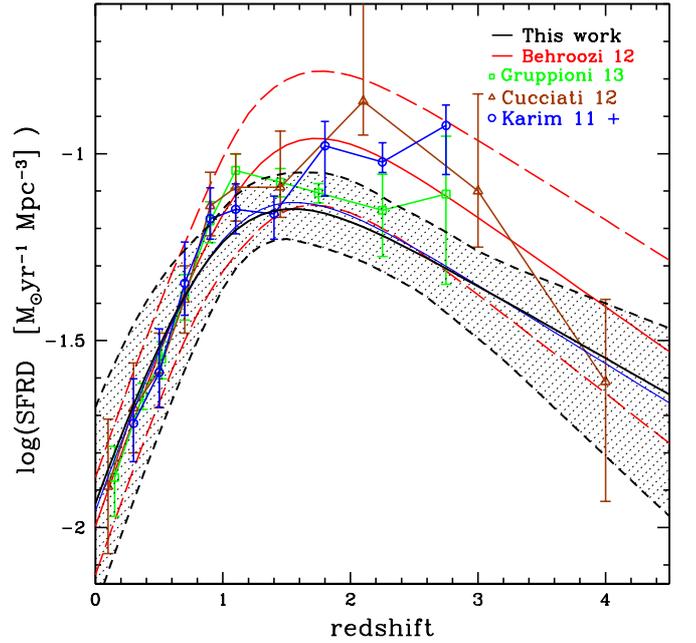}
\caption { A comparison between the star formation history inferred
  from the UltraVISTA mass density (black solid line and dashed area
  corresponding to 1$\sigma$ errors) and literature determinations
  including: direct measurements of the SFR density compiled by
  Behroozi et al. (2013) (red solid line with dashed lines for the
  associated uncertainties); star formation histories derived from the
  UV and IR luminosity functions from Cucciati et al. (2012) and
  Gruppioni et al. (2013) (brown triangles and green squares,
  respectively); and finally radio estimates from Karim et al. (2011)
  updated using the mass functions presented in this paper (blue
  circles). }
           \label{SFRF} 
\end{figure}

\subsection{Inferred star formation history}\label{SFR}

Following Wilkins et al. (2008), we can link the mass density
evolution and the star formation history using
\begin{equation}\label{rhoeq}
\rho_*(t)=\int^t_0 SFRD(t')(1-f_r[t-t'])dt'
\end{equation}
where $SFRD$ corresponds to the star formation rate density and $f_r$
is the stellar mass loss depending on the age of the stellar
populations (Renzini A. \& Buzzoni A., 1986). We adopt the
parametrization of the stellar mass loss provided by Conroy \&
Wechsler (2009) for a Chabrier (2003) IMF
$f_r(t-t')=0.05ln(1+(t-t')/0.3Myr).$

Wilkins et al. (2008) found that the star formation history inferred
from the mass density measurements is not consistent with SFRD
observations. We re-investigate this problem using our own mass
density measurements. We fit the UltraVISTA mass density data using equation
\ref{rhoeq} and the the parameterisation of the star formation
history of Behroozi et al. (2013):
\begin{equation}\label{SFRDpara}
SFRD(z)=\frac{C}{10^{A(z-z_0)}+10^{B(z-z_0)}}.
\end{equation}
The resulting best fit parameters are $B=0.194^{+0.128}_{-0.082}$,
$C=0.111^{+0.040}_{-0.029}$ and $z_0=0.950^{+0.343}_{-0.410}$. We set
$A=-1$ as Behroozi et al. (2013). Our inferred star formation history
and the associated uncertainties are shown with the black solid line
and the shaded area in Figure \ref{SFRF}. The inferred star formation
history is compared with the data compiled by Behroozi et al. (2013)
and the most recent measurements of the SFRD at $0<z<4$ based on UV
(Cucciati et al. 2012), IR (Gruppioni et al. 2013) and radio data
(Karim et al. 2011)\footnote{We updated the values of Karim et
  al. (2011) by using our own MF rather than the ones derived by
  Ilbert et al. (2010)}.  This inferred star formation history is in
excellent agreement with SFRD measurements at $z<1.5$, while we find
differences of 0.2 dex at $z>1.5$. However, such offset is well within
the expected SFRD uncertainties (see the large scatter between the
various SFRD measurements at $z>2$ depending on the wavelength used to trace
the star formation rate in Figure \ref{SFRF}) and the mass density
uncertainties created by the slope extrapolation at the low mass end
(e.g. Mortlock et al. 2011, Santini et al. 2012).

We also derive the mass density evolution expected from the star formation
history compilation of Behroozi et al. (solid green line and shaded
area in Figure \ref{LD}). We find that the expected mass density is
systematically higher by 0.05-0.2 dex than our data, while still
consistent with the expected uncertainties. The discrepancy between
direct and inferred mass densities reaches 0.2 dex at $z\sim 1.5$, and
decreases at lower redshift. We note that we would not observe this
decrease at $z<1.5$ using a constant return fraction.

\begin{figure*}
\centering
\includegraphics[width=16cm]{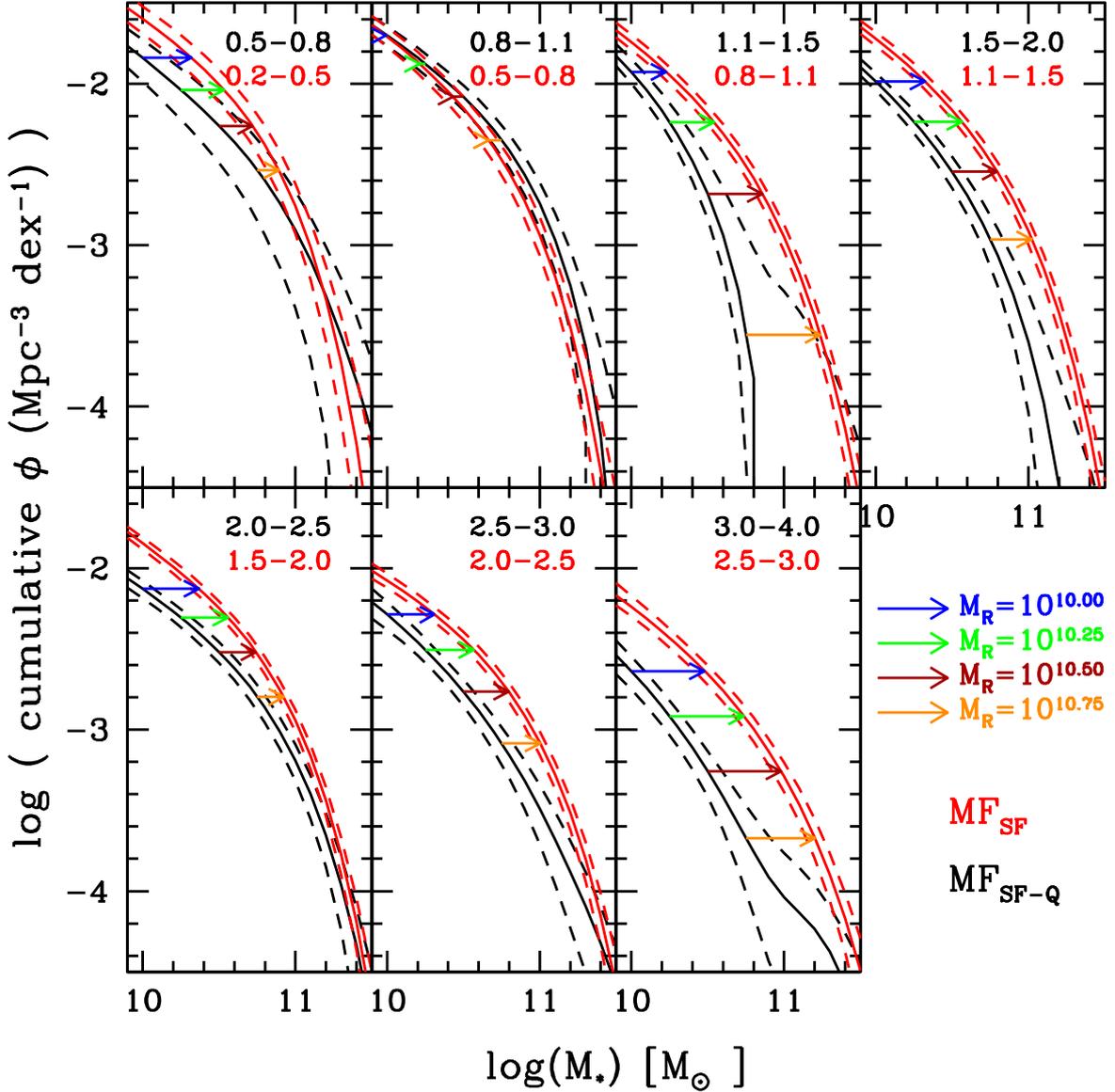}
\caption{In each panel, we show the cumulative MF of star-forming
    galaxies at $t_2$ in red (lowest redshift bin) and at $t_1$ in
    black (highest redshift bin). The contribution of galaxies
    quenched between $t_1$ and $t_2$ is removed from the cumulative MF
    at $t_1$. The dashed lines correspond to the uncertainties. The
    shifts $\Delta \mathrm{log}{\cal M}$ used to estimate the sSFR are
    shown with the horizontal arrows at four reference masses: ${\cal
      M}_R=10^{10}$, $10^{10.25}$, $10^{10.5}$, $10^{10.75}{\cal
      M}_{\sun}$ in blue, green, brown and orange, respectively.}
           \label{LFcumu} 
\end{figure*}

\begin{figure}
\centering
\includegraphics[width=9.cm]{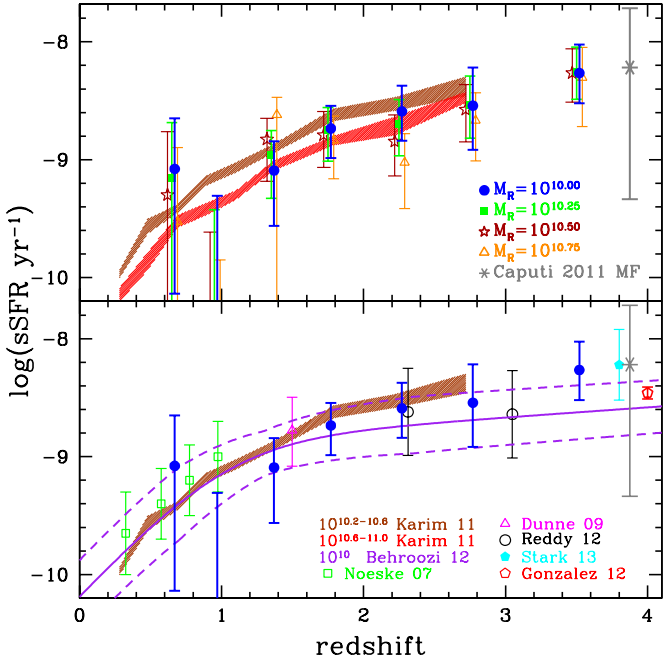}
\caption{Evolution of the sSFR (SFR/${\cal M}$) derived from the
  UltraVISTA mass functions. {\it Top panel:} sSFR measured at
  different masses using the shifts $\Delta \mathrm{log}{\cal M}$
  shown in Figure \ref{LFcumu}. The sSFR computed at four reference
  masses ${\cal M}_R=10^{10}$, $10^{10.25}$, $10^{10.5}$,
  $10^{10.75}{\cal M}_{\sun}$ are shown with blue circles, green
  squares, brown stars and orange triangles, respectively. We also
  added with a grey star the sSFR measured using the same method using
  the MF of Caputi et al. (2011). The brown and red shaded areas for
  Karim et al. (2011) correspond to the mass bins $10^{10.2}-10^{10.6}
  {\cal M}_{\sun}$ and $10^{10.6}-10^{11} {\cal M}_{\sun}$,
  respectively. {\it Bottom panel:} sSFR estimated at $10^{10}{\cal
    M}_{\sun}$ compared with direct measurements from the
  literature. The purple solid line (dashed lines showing the
  uncertainties) correspond to the compilation by Behroozi et
  al. (2013) at the mass $10^{10} {\cal M}_{\sun}$, respectively. The
  other points are a compilation by Stark et al. (2013) including also
  data from Dunne et al. (2009) and Reddy et al. (2012) at $log({\cal
    M}) \sim 9.7$. The red pentagon corresponds to the Gonzalez et
  al. (2012) value at $log({\cal M}) \sim 9.7$. The open green squares
  correspond to Noeske et al. (2007) at $log({\cal M}) \sim 10$.}
           \label{ssfrF} 
\end{figure}

\subsection{Inferred Specific Star Formation Rates}\label{ssfr}

We now consider an admittedly over-simplistic scenario in which
evolution of star-forming galaxies is driven only by star formation
(i.e. we consider that mergers do not significantly change the galaxy
distribution between two redshift bins, see Section 4.3 of Boissier et
al. 2010). Given this assumption, the stellar masses increase by 
  ${\cal M}(t_2)-{\cal M}(t_1)=\int^{t_2}_{t_1}
  SFR(t')(1-f_r[t_2-t'])dt'$ between $t_1$ and $t_2$ ($t_1<t_2$), with
  $f_r$ corresponding to the stellar mass loss (see section
  \ref{SFR}). Assuming that the SFR remains constant over the
  considered time interval and over the mass range $[{\cal
    M}(t_1),{\cal M}(t_2)]$, we obtain the specific SFR:
\begin{equation}\label{ssfreq}
\mathrm{sSFR}(t_1)=\frac{10^{\Delta \mathrm{log}{\cal M}}-1}{(t_2-t_1-\int^{t_2}_{t_1}f_r(t_2-t')dt')}
\end{equation}
with $\Delta \mathrm{log}{\cal M}=\mathrm{log}{\cal
  M}(t_2)-\mathrm{log}{\cal M}(t_1)$. The shift $\Delta
\mathrm{log}{\cal M}$ is directly derived from the MF evolution of
star forming galaxies between $t_1$ and $t_2$.

Star-forming galaxies could be quenched and move to the quiescent
population in the time interval $t_2-t_1$. Since we want to compute
$\Delta \mathrm{log}{\cal M}$ for the same galaxy population at $t_1$
and $t_2$, we need to remove the contribution of the galaxies quenched
between $t_1$ and $t_2$. This contribution is simply the difference
between the quiescent MF (hereafter ${\rm MF}_{\rm Q}$) estimated at
$t_2$ and $t_1$. Therefore, the MF of star-forming galaxies without
the contribution of the galaxies quenched between $t_1$ and $t_2$ is
computed as ${\rm MF}_{\rm SF-Q}(t_1)={\rm MF}_{\rm SF}(t_1)-{\rm
  MF}_{\rm Q}(t_2)+{\rm MF}_{\rm Q}(t_1)$.

We measure the shifts $\Delta \mathrm{log}{\cal M}$ required to
superimpose the cumulative MF$_{\rm SF-Q}(t_1)$ and of the cumulative
MF$_{\rm SF}(t_2)$. These shifts are indicated with horizontal arrows
in Figure \ref{LFcumu}. Since the sSFR could depend on the mass
(e.g. Dune et al. 2009, Karim et al. 2011), we measure $\Delta
\mathrm{log}{\cal M}$ at four reference masses (${\cal M}_R=10^{10}$,
$10^{10.25}$, $10^{10.5}$ and $10^{10.75}{\cal M}_{\sun}$). We do not
consider ${\cal M}_R<10^{10}{\cal M}_{\sun}$ to limit the impact of
the slope extrapolation at low masses. The top label of Figure
\ref{ssfrF} shows the evolution of the sSFR estimated at the four
reference masses in our analysis. We find consistent sSFR for the
three reference masses lower than ${\cal M}< 10^{10.75}{\cal
  M}_{\sun}$. In the bottom panel of Figure \ref{ssfrF}, we focus on
the sSFR evolution measured for ${\cal M}_R=10^{10}{\cal M}_{\sun}$,
which depends less on the removal of quenched galaxies. The sSFR
increases from $z=1$ to $z=4$ (blue circles). The sSFR computed with
this indirect method are compared with direct measurement of the sSFR
from the literature.  Given the size of the uncertainties, our
inferred sSFR is in good agreement with literature measurements. At
$z>2.5$, we obtain a sSFR larger than the compilation from Behroozi et
al. (2013). However, recent studies taking into account the
contribution of nebular emission lines show that the sSFR at $z>2$
could be higher than previously found (e.g. de Barros et al. 2012,
Stark et al. 2013). Still, we obtain a value at $z\sim 3.5$ which is
higher than Gonzalez et al. (2012) which takes nebular emission lines
into account. We also apply the same method with the MFs computed by
Caputi et al. (2011) to derive the sSFR at $z \sim 4$ (grey star in
Figure \ref{ssfr}). Given the mass limit of this study, we adopt a
reference mass of ${\cal M}_R=10^{10.5}{\cal M}_{\sun}$. Despite large
error bars, our sSFR values are in excellent agreement.

Finally, we note that our method is more robust at $z>1.5$ than at
lower redshift, as shown by the size of the error bars. Indeed, the
contribution of the quiescent galaxies to the global population is
below 40\% at $z>1.5$ and is restricted to the massive population
(bottom panel of Figure \ref{MFtype}). Therefore, residual
uncertainties on the removal of the quenched population has a small
impact on the analysis. At $z<1.5$, the quiescent galaxies are built
very fast and their contributions to the global MF reaches 80\% at
$0.8<z<1.1$. Therefore, the uncertainties on $MF_Q$ generated by
cosmic variance have a strong impact on our density estimate of the
quenched population between $t_1$ and $t_2$.

\begin{figure*}
\centering
\includegraphics[width=11cm,angle=90,trim = 35mm 0mm 0mm 0mm]{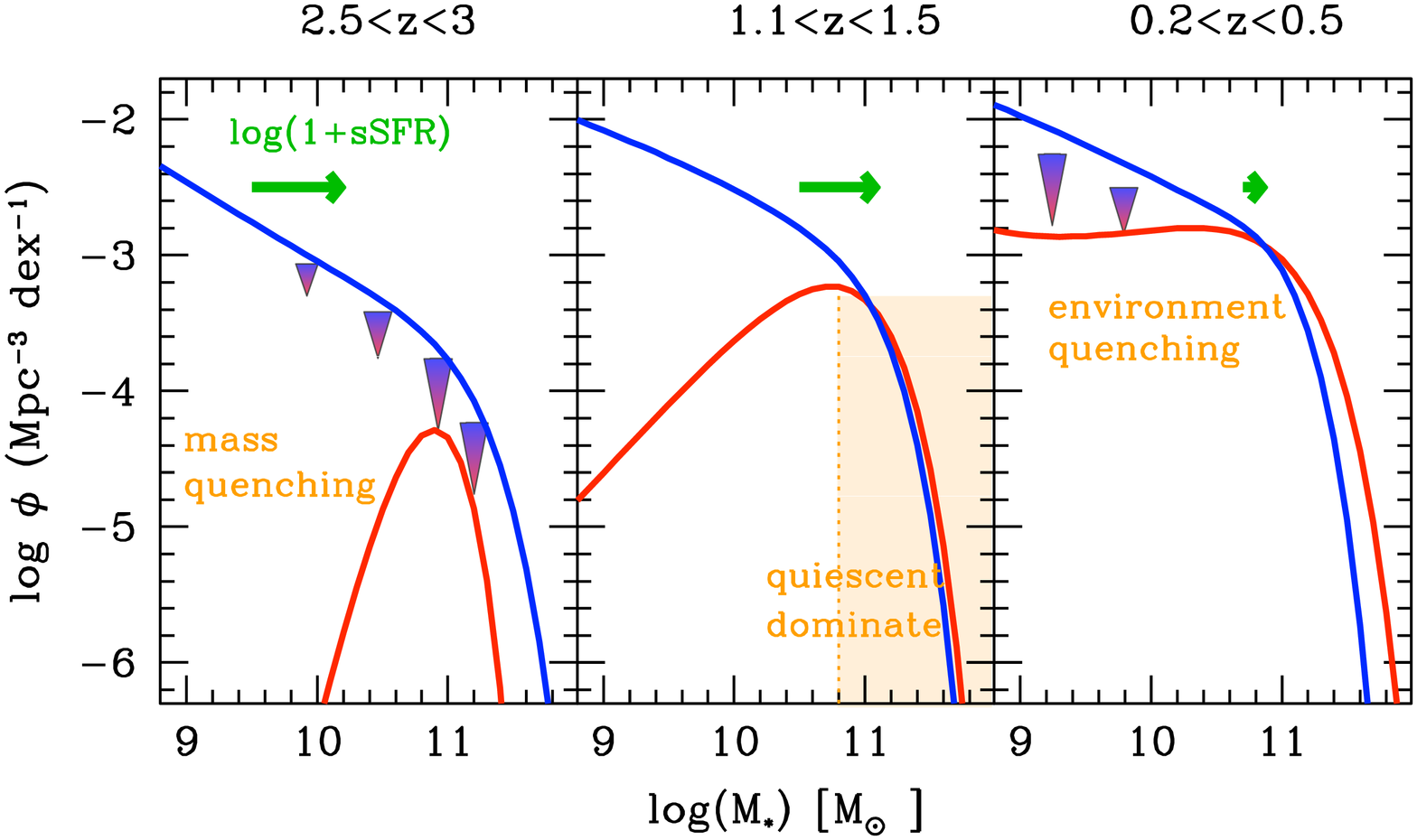}
\caption{A possible scenario showing how sSFR and quenching impact the
  star-forming MF (blue lines) and quiescent MF (red lines), similar
  to Peng et al. (2010).  The large arrows represent quenching. The
  green arrows correspond to the mass increase expected in 2 Gyr, by
  taking the sSFR values compiled by Stark et al. (2013).  The left
  panel corresponds to the high redshift bin $2.5<z<3$, where we show
  the mass quenching process which is more efficient at high mass. The
  middle panel corresponds to the redshift bin $1.1<z<1.5$ when the
  quiescent galaxies starts to dominate the high-mass end. The right panel
  corresponds to the redshift bin $0.2<z<0.5$, when environment
  quenching generates new low mass quiescent galaxies.}
           \label{schema}%
 \end{figure*}

\subsection{Quenching processes}\label{quenching}

Newly quenched galaxies and major mergers drive the evolution in
density of the quiescent MF, since by definition a quiescent galaxy
grows by less than 0.03 dex in 6 Gyr by star formation. Therefore, the
efficiency of quenching with cosmic time can be directly estimated
from the MF evolution of the quiescent and star-forming galaxies. In
this section we discuss our results in term of quenching efficiency,
and we summarize our interpretation with a synthetic scheme (Figure
\ref{schema}).

As we discuss in \S\ref{ssfr}, the evolution of the low-mass
star-forming galaxies at $z>2$ is consistent with the expected sSFR
while the lack of evolution of the massive end implies that massive
star-forming galaxies are necessarily quenched. Our data are deep
enough at $z<3$ to show that the slope of the star-forming sample is
steeper than the slope of the quiescent population (see Figure
\ref{MFtype}). If the quenching processes acted with the same
efficiency at all masses, the MF of the quiescent population should
have the same slope as the star-forming MF. We conclude that the
physical process which quenches star formation is necessarily more
efficient above ${\cal M} \gtrsim 10^{10.7-10.9} {\cal M}_{\sun}$,
i.e. the maximum in density of the quiescent MF\footnote{The maximum
  in density of $\phi( \mathrm{log}{\cal M})$ occurs at $
  \mathrm{log}({\cal M}^*)+ \mathrm{log}(\alpha+1)$}. This scenario is
fully consistent with the model proposed in Peng et al. (2010) who
introduce a ``mass quenching'' process which is independent of
environment for the massive galaxies. We show that this mass quenching
has been active since $z=3$. This mass quenching process must have
been efficient at $1<z<3$ to rapidly build up the quiescent population
(left panel of Figure \ref{schema}). For instance, around 30\% of the
star-forming galaxies with ${\cal M}\sim 10^{11} {\cal M}_{\sun}$
should be quenched between $2.5<z<3$ and $2<z<2.5$ to explain a
density increase by 0.25 dex of massive quiescent galaxies. While we
can see that this quenching process is extremely efficient above a
given mass, we cannot infer the physical nature of the quenching
process. AGN feedback or major mergers (or both) are possible
candidates (e.g. Hopkins et al. 2006).

Since the star-forming galaxies dominate the high-mass end of the
global MF at $z>1$ (bottom panel of Figure \ref{MFtype}), an efficient
``mass quenching'' process is rapidly transitioning them into passive
systems. If the ``mass quenching'' process depletes more rapidly the
reservoir of massive star-forming galaxies than new star formation is
able to replenish it, we reach an epoch where the quenching is no
longer a channel to create new massive quiescent galaxies.  By $z\sim
1$ the supply of massive star forming galaxies has dwindled to the
point that few new galaxies are being quenched. Figure \ref{MFone}
shows that $>70$\% of the galaxies more massive than ${\cal M}>10^{11}
{\cal M}_{\sun}$ are quiescent at $z<1.1$. Moreover, the sSFR
estimated from the literature decreases by a factor 15 between $z\sim
1.5$ and $z\sim 0.6$ as shown by the green arrows in Figure
\ref{schema} (e.g. Noeske et al. 2007, Daddi et al. 2007, Elbaz et
al. 2007, Karim et al. 2011). Since the massive galaxy population is
already dominated by quiescent galaxies at $z<1$ and since the growth
rate of star-forming galaxies is no longer sufficiently efficient to
generate numerous massive star-forming galaxies (middle panel of
Figure \ref{schema}), the quenching of massive star-forming galaxies
can not modify significantly the high-mass end of the quiescent MF at
$z<1$.

The flattening of the faint-end slope of the quiescent MF at $z<1$
(see \S\ref{ell}) shows that the quenching starts to be efficient at
low masses (right panel of Figure \ref{schema}). As shown by Bezanson
et al. (2012), such flattening of the slope is easily obtained by
quenching a small fraction (between 1 and 10\%) of the star-forming
galaxies. Using the COSMOS sample, Scoville et al. (2013) find that
low-mass quiescent galaxies appear in high density environment at
$z<1$. Peng et al. (2010) proposed the ``environment quenching''
process which mainly affects the satellite galaxies as large-scale
structure develop. Gabor \& Dav\'e (2012) find a similar quenching
process using an hydronamical simulation. Therefore, we interpret the
flattening of the slope of the quiescent MF as an environmentally
driven process, even if our results alone can not demonstrate it.

\subsection{Impact of the assumed star formation histories}\label{sfrhist}

We derive the stellar masses assuming an exponentially declining star
formation rate for the stellar population synthesis models. However,
numerous studies pointed out that such star formation histories could
be inappropriate at $z>1$ (e.g. Maraston et al. 2010, Pacifici et
al. 2013). Therefore, we investigate how this choice impacts our
conclusions.

We derive the MF using two other libraries: one library based on {\it
  delayed} star formation histories and another using a {\it
  composite} of various star formation histories (see
\S\ref{stellarmasses}). We generate the second library based on
delayed star formation histories following a law in $\tau^{-2} t
e^{-t/\tau}$. The maximum SFR peak
could be delayed by 0.1, 0.5, 1, 2 or 3 Gyr and we consider three
different metallicities ($Z=0.004$, $Z=0.008$, $Z=0.02$). We generate
a third {\it composite} library including various combinations: three
templates with an exponentially declining SFR ($\tau=0.1, 1, 5$ Gyr),
two templates with a delayed star formation history (maximum peak at 1
and 3 Gyr). For each of these templates, we allow a possible second
burst 3 Gyr after the peak in SFR. The amplitude of this second burst
could be 0.25 or 0.5 times the maximum of the first burst. This second
burst is generated using a constant SFR. We derive the MFs with these
three libraries (called hereafter {\it exponential}, {\it delayed} and
{\it composite} libraries). Still, we do not test the impact of
  stochastic star formation histories.

Figures \ref{MFsimuSF} and \ref{MFsimuEll} show the MFs of the
star-forming and quiescent galaxies, respectively. The MFs based on
the {\it exponential}, {\it delayed} and {\it composite} libraries are
shown with red squares, blue triangles and green crosses,
respectively. We do not detect any significant difference between the
MFs. Therefore, all our conclusions are equally valid for any of these
three libraries.

In detail, we find that quiescent galaxies are almost always fitted
with a {\it delayed} SFR history peaking after 0.1 or 0.5 Gyr, which
is close to an exponentially declining history. For the star forming
galaxies, the templates are uniformly distributed between the five
{\it delayed} star formation histories. By comparing the best-fit ages
and the $\tau$ values of the delayed templates, 30\% of star-forming
galaxies at $z>2$ are fitted with a template in the rising part of the
star formation history. The dispersion between the stellar masses
computed with the {\it exponential} and {\it delayed} libraries is
lower than 0.06 dex out to $z<4$, with no systematic difference. This
explains why the {\it delayed} and {\it exponential} MFs are
similar. For the {\it composite} library, the templates with an
exponentially declining star formation history are selected in 87\% of
the case, with a possible small second burst. This explains why we
obtain almost the same results between the {\it composite} and the
{\it exponential} libraries.

\subsection{Comparison with the semi-analytical models}

We now compare our mass function measurements with the predictions of
semi-analytical models. The mock catalogs are based on $\Lambda$CDM
simulations from Wang et al. (2008) with the cosmological parameters
derived from the third-year WMAP data ($H_0=74.3$ km/s,
$\Omega_M=0.226$ and $\Omega_{\Lambda}=0.774$). The light cone survey
covers an area of $1.4\times 1.4$ deg$^2$ similar to COSMOS. Galaxy
properties were generated using the galaxy formation model, as
detailed in De Lucia \& Blaizot (2007) and Wang \& White (2008). Since
the redshift and the galaxy stellar masses are available for all
galaxies in the simulation, we can directly compute the predicted
MFs. We use our standard cosmology ($H_0=70$ km/s, $\Omega_M=0.3$ and
$\Omega_{\Lambda}=0.7$) to renormalise the predicted stellar mass
counts by the comoving volume. We isolate the quiescent population in
the simulations. We try three different criteria: a red clump criteria
using rest-frame $UV-r$ versus $r-K$ colours, a cut in $log(sSFR)<-11$
and a cut in $B-I$ corrected for dust extinction. The quiescent MFs
are almost identical regardless of selection criteria.

\begin{figure*}
\centering
\includegraphics[width=16cm]{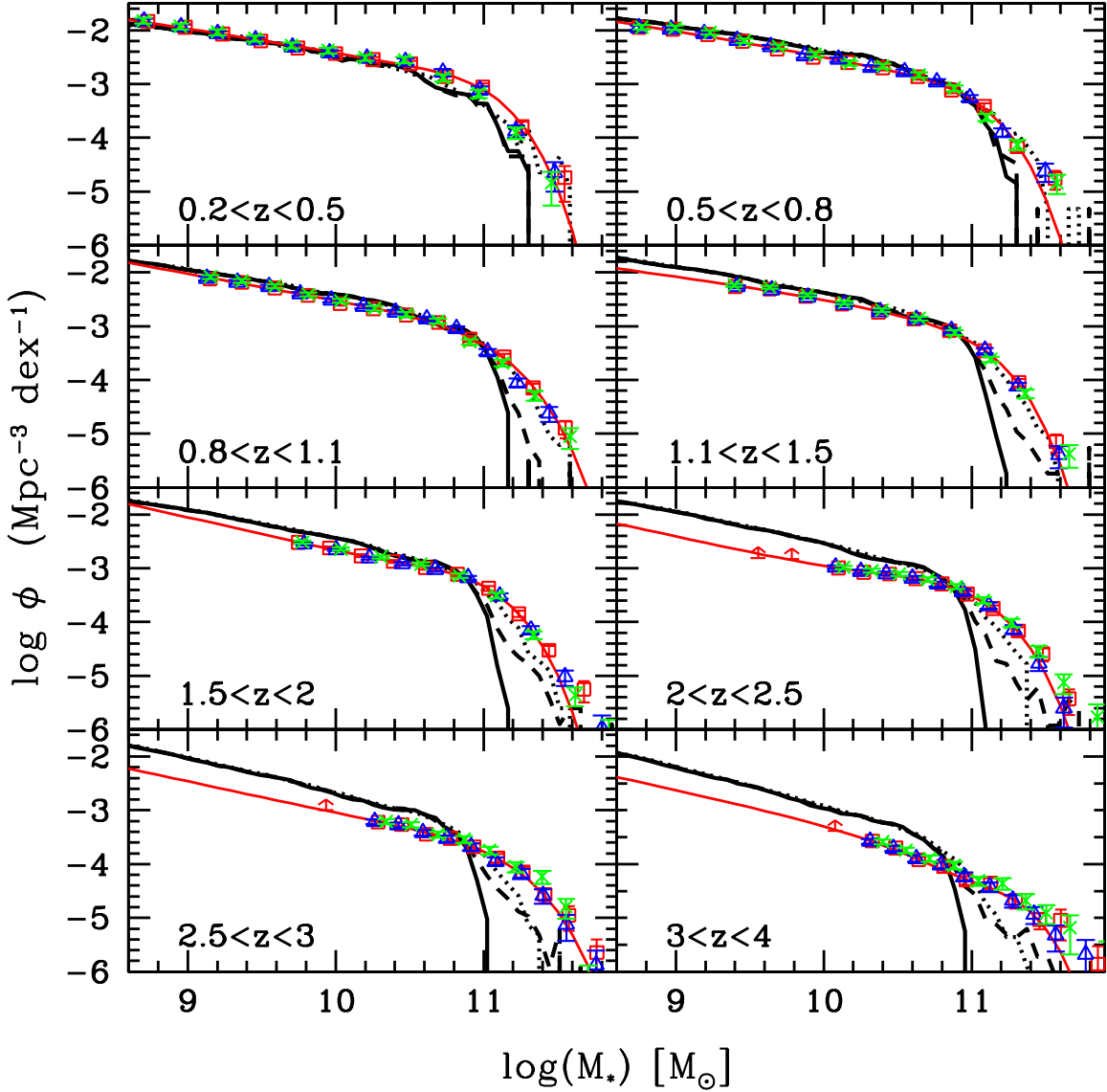}
\caption{Galaxy stellar mass functions of the star-forming sample. The
  points correspond to the {\it Vmax} estimator (we do not show the
  other estimators for clarity) and error bars include cosmic
  variance, Poissonian errors and photo-z/stellar mass uncertainties.
  The red squares, the blue triangles and the green crosses correspond
  to the MFs computed using the {\it exponential}, {\it delayed} and
  {\it composite} libraries. The red solid line is the best-fitted
  double-Schechter function. The black solid, dashed and dotted lines
  correspond to the MFs predicted by the semi-analytical model,
  assuming no error on the stellar masses, an error evolving as
  described in appendix A (using the product of a Lorentzian and
  gaussian distribution), and a gaussian with a $\sigma=0.2 dex$. }
           \label{MFsimuSF}%
\end{figure*}

The comparison between the observed and predicted MFs is shown in
Figure \ref{MFsimuSF} for the star-forming sample. First, we find that
the predicted and observed faint-end slopes are in good agreement at
least up to $z=2$. Secondly, the model under-predicts the density of
massive galaxies. The mismatch between the observed and predicted
high-mass end increases with redshift. If we add a 1$\sigma$ error of
0.2 dex to the stellar masses, as commonly assumed in the literature
(e.g. Cattaneo et al. 2006, Bower et al. 2012), we reduce the tension
between model and observation (dotted lines in Figure
\ref{MFsimuSF}). We point out that the 0.2 dex uncertainty commonly
associated to the stellar mass is mostly explained by systematic
uncertainties (choice of the IMF, adopted population synthesis models)
and should not be applied to the stellar masses. For a given library,
the statistical errors are much smaller. Moreover, we already
deconvolved the fit by the expected statistical errors in the stellar
masses (see \S\ref{estimate}). Therefore, the red solid line is
directly comparable to the model prediction, showing that the
predicted high-mass end is underestimated. Several factors could
explain this mismatch: 1) we do not consider stochastic star
  formation histories to estimate the stellar masses, while such
  histories are common for simulated galaxies in semi-analytical
  models. Such difference could create an intrinsic scatter which
  could justify a Gaussian smoothing of the observed MF; 2) our high
redshift sample is contaminated by low redshift galaxies
(i.e. catastrophic failures in the photo-z); however Figure
\ref{zp_zs} do not show such contamination with our current
spectroscopic sample; 3) some assumed physical laws (e.g. IMF) evolve
with redshift which generate redshift dependent biases in our stellar
mass estimates; 4) the models do not generate enough massive galaxies
at $z=3-4$. The last option would require that the models generate
$10^{11.5}{\cal M}_{\sun}$ galaxies in less than two Gyr. Then, these
galaxies should stop to grow on a very short timescale by having their
star formation activity drastically reduced or quenched.

\begin{figure*}
\centering
\includegraphics[width=16cm]{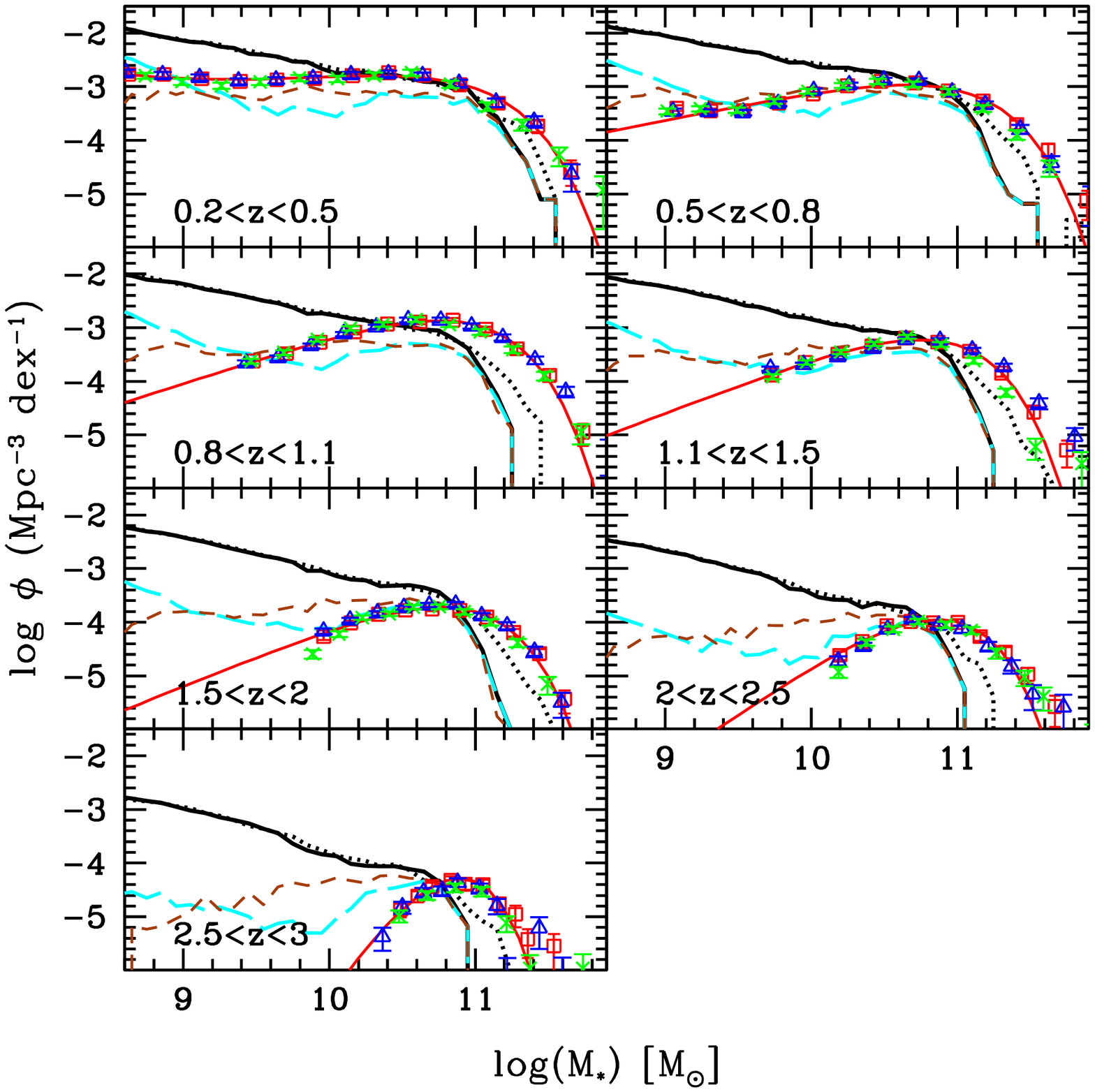}
\caption{Galaxy stellar mass functions of the quiescent sample for
  selected using a two-colour technique. Symbols are the same as Figure
  \ref{MFsimuSF}. The short dashed brown lines correspond to quiescent
  galaxies with a bulge-to-disk ratio greater than 0.5. The long
  dashed cyan lines correspond to the central galaxies in the models.}
           \label{MFsimuEll}%
\end{figure*}

Figure \ref{MFsimuEll} shows the comparison between observed and
predicted MFs for the quiescent population. The mismatch between the
observed and predicted faint end slopes is dramatic. At $9.5<log({\cal
  M})<10$ and $0.5<z<0.8$, the model overestimates the low-mass end by
a factor 10. Such an effect has been already noticed by Cucciati et
al. (2012) in the VVDS field based on the B-band luminosity function.
Wang \& White (2008), Bielby et al. (2012), Bower et al. (2012), Guo
et al. (2011) show that the models overproduce the density of low-mass
galaxies for the full population. In this paper, we show that the
origin of this discrepancy is the high density of low-mass quiescent
galaxies, since the faint end slope of the star-forming galaxies is
correctly predicted at $z<2$. Information such as the host halo mass,
the galaxy location in the halo (central or satellite) and the galaxy
morphology are kept in the simulation. We use this information to
investigate in more detail the properties of the quiescent galaxies
overproduced by the models. We first keep only bulge-dominated
quiescent galaxies with a bulge-to-disk ratio greater than 0.5 (brown
short dashed lines). The slopes are in much better agreement with the
data.  Therefore, the model is able to generate a reasonable amount of
quiescent galaxies when the quenching is associated to a process which
makes the galaxy dominated by a bulge. Indeed, Ilbert et al. (2010)
showed that between 40 and 70\% of the quiescent galaxies have an
elliptical morphology at $z<0.8$ and $9.5<log({\cal M})<10$, while in
the model, $20$\% of the quiescent galaxies are bulge dominated in the
same mass range. We also show the predicted MF of central quiescent
galaxies (cyan long dashed lines). It appears that the low-mass
quiescent simulated galaxies are mostly dominated by satellite/orphans
galaxies. One possible interpretation of the mismatch between the
faint-end slopes is an over-quenching of the star formation in
satellite galaxies and/or an over-quenching of the star-formation in
disk galaxies.

\section{Conclusions}

We study the stellar mass assembly out to $z=4$ with a unique sample
of 220,000 galaxies selected at $K_{\rm s}<24$ in the COSMOS field.
Galaxies are selected using the new UltraVISTA DR1 near-infrared data
release, which reaches between one and two magnitudes deeper than
previous COSMOS near-infrared data. Fluxes are measured on
PSF-homogenised images in 25 bands including optical broad and
intermediate-band Subaru data taken for the COSMOS project. We also
add four bands at $3.6-8\mu m$ from S-COSMOS and the GALEX NUV
band. We keep only the sources at $K_{\rm s}<24$ located in areas with
a good image quality, which represents an effective area of 1.52
deg$^2$.

The photometric redshifts are derived using \texttt{Le\_Phare}
following a procedure similar to Ilbert et al. (2009). Our photometric
redshifts have two regimes: at $i^+_{AB}<22.5$ ($z_{med}\sim 0.5$),
precision $1\%$ with less than 1\% of catastrophic failures. In the
high redshift range $1.5<z<4$, the precision of the photo-z is
tested against the zCOSMOS faint sample, and is 3\% for $i^+_{med}
\sim 24$ galaxies. We estimate the galaxy stellar masses using a
library of synthetic spectra generated using the SPS model of Bruzual
and Charlot (2003). We show that our results are independent of the
assumed star-formation history (among our tested star-formation
  histories, i.e. exponentially decreasing, delayed or a possible
  second burst).

We find that the evolution of the global stellar mass function is
strongly mass-dependent. The low-mass end evolves more rapidly than
the high-mass end. If we consider an evolution purely driven by star
formation, the low-mass galaxies evolve by almost 1 dex between
$1.5<z<2$ and $0.2<z<0.5$, while the stellar mass of the most massive
galaxies increases by less than 0.2 dex in the same time interval. The
lack of evolution of the massive end can be interpreted as a direct
consequence of star formation being drastically reduced or quenched
when a galaxy becomes more massive than ${\cal M}>10^{10.7-10.9} {\cal
  M}_{\sun}$. We derive the comoving stellar mass density of the full
sample from $z=4$ to $z=0.2$. We find that the stellar mass is
assembled twice as fast at $1<z<4$ (14\% of the local stellar mass
density per Gyr) than at $z<1$ (6\% of the local stellar mass density
per Gyr).

We isolate the quiescent population using a classification based on
the rest-frame colours $NUV-r^+$ and $r^+-J$. This classification
separates cleanly dusty star-forming galaxies and quiescent
galaxies. We find that the MF evolution of the quiescent population is
definitively mass-dependent at $z<1$, confirming the trend shown in
Ilbert et al. (2010). We do not find any significant evolution of the
high-mass end at $z<1$ (any evolution being limited to $\Delta M < 0.2
dex$), while we observe a clear flattening of the slope in the same
redshift range. We interpret this evolution of the low-mass end of the
MF as arising from continuous quenching of galaxies between $z\sim 1$
and $z\sim 0.1$, probably by physical processes related to 
environment.

The UltraVISTA data now allows us to trace the growth in stellar mass
density in quiescent galaxies from $z\sim3$ to the present day. From
$z\sim3$ to $z\sim1$ we find a rapid increase in the stellar mass
density of all quiescent galaxies, independent of stellar mass. We
confirm that the steep rise of more than one order of magnitude of
this population between $1<z<2$ observed in previous works (Arnouts et
al., 2007; Ilbert et al., 2010) follows an earlier rise by a factor
$\sim3$ in $2<z<3.5$. In total, we find that the density of $10^{11}
{\cal M}_{\sun}$ galaxies is continuously increasing, reaching a
factor $\sim40$ between $2.5<z<3$ and $0.8<z<1.1$. This must indicate
that a fraction of star-forming galaxies is continuously quenched at
$z>1$. Because of the different faint-end slopes of the MF between the
quiescent and global populations at $z<3$, we infer that the physical
processes which quench the star formation are more efficient above
${\cal M} \gtrsim 10^{10.7-10.9} {\cal M}_{\sun}$ (the maximum in density
of the quiescent MF). This scenario is consistent with the model
proposed by Peng et al. (2010) who introduce a ``mass quenching''
process. Since the high-mass end of the quiescent MF stops evolving at
$z<1$, we conclude that: 1) star formation is not efficient enough at
$z<1$ to produce new massive star-forming galaxies, which could be
quenched later; 2) major mergers between massive galaxies are not
sufficiently frequent at $z<1$ to increase significantly the density
of massive quiescent galaxies.

We infer the star formation history from the mass density evolution
following the formalism of Wilkins et al. (2008). We find that the
inferred star formation history is in excellent agreement with SFRD
data at $z<1.5$ but under-predict them by 0.2 dex at $z>1.5$. However,
considering the size of the uncertainties at $z>1.5$, the SFRD and
mass density data still provide a consistent picture of the star
formation history.

We also develop a new method to estimate the sSFR from the mass
function evolution. By comparing the star-forming MFs at two different
epochs (after having removed the contribution of galaxies quenched
during the considered time interval), we derive the sSFR from $z \sim
0.5$ up to $z<4$.  We find that the sSFR increases continuously at
$1<z<4$ for our considered mass range $10^{10}{\cal M}_{\sun} \leq
{\cal M} \leq 10^{10.5}{\cal M}_{\sun}$. This new method is
complementary to direct sSFR measurements at $z>2$, which are very
sensitive to dust extinction (e.g. Bouwens et al. 2009) or SED
modeling (e.g. Stark et al. 2013).

Finally, we compare our data with the predictions of the
semi-analytical model of De Lucia and Blaizot (2007) as described in
Wang et al. (2008).  We find that the model under-predicts the density
of massive galaxies for quiescent and star-forming galaxies. The
tension between model and observations at the high-mass end can be
alleviated if scatter in the modeled stellar masses errors is a
1$\sigma$ of 0.2 dex. When we compare the observed and predicted MFs
for the quiescent population, the difference in the slope is
striking. At $log({\cal M})\sim 10$ and $z<0.8$, the model
overestimates the low-mass end by a factor 10. On the other hand, the
model successfully reproduce the faint-end slope of the star-forming
population. This indicates that recipes to quench satellite and/or
disk galaxies within semi-analytical models are still incomplete and
will need to be modified.

The COSMOS field will be observed during five years for the UltraVISTA
survey, allowing to obtain NIR data two magnitudes deeper. In the
meantime, the SPLASH program (Capak et al., in preparation) will
observe the COSMOS field with IRAC, allowing a gain of 1.5 mag at
$3.6\mu m$. The combination of this new data will allowed to extend
this work at $4<z<7$ and better constrain the contribution of low-mass
galaxies.

\begin{acknowledgements}
We are grateful to the referee for a careful reading of the
manuscript and useful suggestions. The authors thank
M. Kajisawa, L. Pozzetti, P. P\'erez-Gonz\'alez and P. Santini for
providing their estimates of the mass function. We also thanks
P. Behroozi for his useful suggestions. The authors thank J. Donley
for providing her catalogue of IRAC power-laws. We gratefully
acknowledge the contributions of the entire COSMOS collaboration
consisting of more than 100 scientists.  The {\it HST} COSMOS program
was supported through NASA grant HST-GO-09822.  More information on
the COSMOS survey is available at
http://www.astro.caltech.edu/cosmos. H.J. McCracken acknowledges
support from the `Programme national cosmologie et galaxies''.  JSD
acknowledges the support of the European Research Council through an
Advanced grant, and the support of the Royal Society via a Wolfson
Research Merit Award. S. Toft acknowledges support from the Lundbeck
foundation. The Dark Cosmology Centre is funded by the Danish National
Research Foundation. This paper is based on observations made with
ESO Telescopes at the La Silla Paranal Observatory under ESO programme
ID 179.A-2005 and on data products produced by TERAPIX and the
Cambridge Astronomy Survey Unit on behalf of the UltraVISTA
consortium. This work is based in part on archival SEDS data obtained
with the Spitzer Space Telescope, which is operated by the Jet
Propulsion Laboratory, California Institute of Technology under a
contract with NASA. Support for this work was provided by NASA. This
research has made use of the NASA/ IPAC Infrared Science Archive,
which is operated by the Jet Propulsion Laboratory, California
Institute of Technology, under contract with the National Aeronautics
and Space Administration. The Dark Cosmology Centre is funded by the
Danish National Research Foundation.  BMJ acknowledges support from
the ERC-StG grant EGGS-278202.
\end{acknowledgements}

\appendix

\section{Treatment of the Eddington bias}\label{eddi}

Caputi et al. (2011) outlined that a bias (Eddington 1913) could
affect the estimate of the high-mass end of the stellar mass function.
The uncertainties on the stellar masses scatter the galaxies from one
stellar mass bin to another. In the case of the stellar mass function,
the density exponentially decreases toward massive galaxies. The
stellar mass scattering therefore moves more galaxies into the massive
end than the reverse. As a consequence, we overestimate the density of
massive galaxies.

\begin{figure}
\centering \includegraphics[width=9cm]{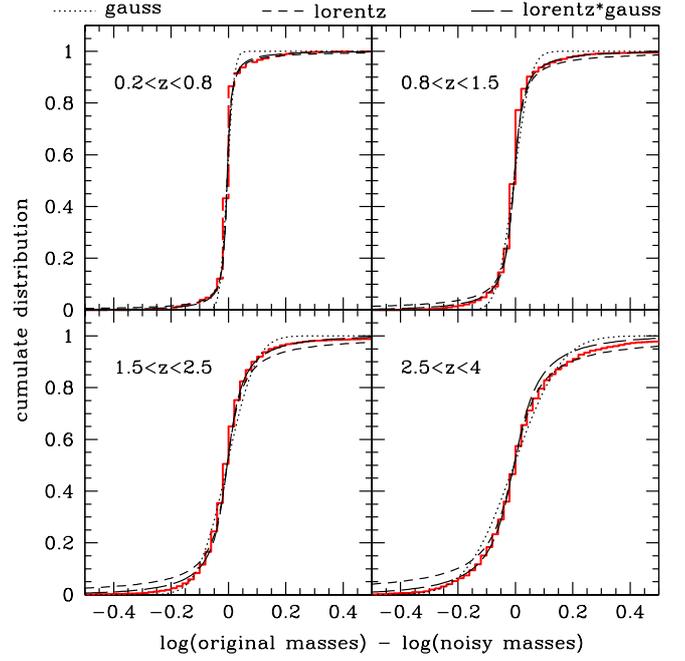}
\caption{The red histogram show the cumulative distribution of the
  difference between the original stellar masses and the ones obtained
  with a ``noisy'' catalogues (redshifts and magnitudes are scattered
  according to the expected errors) at ${\cal M}>10^{10.5}{\cal
    M}_{\sun}$. The dotted, short-dashed and long-dashed lines
  correspond to a Gaussian distribution with $\sigma=0.04(1+z)$, a
  Lorentzian distribution with $\tau=0.04(1+z)$, and the product of a
  Lorentzian and a Gaussian distributions with $\tau=0.04(1+z)$ and
  $\sigma=0.5$.}
           \label{masserr}
\end{figure}

Caputi et al. (2011) estimated that the Eddington bias could affect
the density by up to 0.13 dex for the most massive galaxies. In this
paper, we go one step further and we deconvolve the density estimate
by the stellar mass uncertainties. The first step is to characterise
the stellar mass uncertainties. We do not consider here systematic
errors linked to the choice of the stellar synthesis models or to the
IMF. We consider only the uncertainties linked to the photometry and
to the photometric redshifts. We generate a ``noisy'' sample by adding
noise to the apparent magnitudes and to the photometric redshifts
using their associated error bars. Then, we recompute the stellar
masses with this ``noisy'' catalogue. Figure \ref{masserr} shows the
difference between the original masses and the ones obtained on this
``noisy''catalogue at ${\cal M}>10^{10.5}{\cal M}_{\sun}$. The red
histogram shows the cumulative distribution of these differences.  We
first model the stellar mass uncertainties using a Gaussian
distribution $G(x)=\frac{1}{\sigma \sqrt{2\pi}} exp^{-\frac{1}{2}
  \frac{x}{\sigma}^2}$ as it is commonly assumed (e.g. Cattaneo et
al. 2006, Bower et al. 2012). The dotted lines correspond to a
Gaussian distribution with $\sigma=0.04(1+z)$. The fraction of
galaxies with a difference larger than 0.1 dex is underestimated and
the Gaussian distribution is not a good representation of the main
peak at $z>2.5$. We also model the distribution by using a Lorentzian
distribution $ L(x)=\frac{\tau}{2\pi}\frac{1}{(\frac{\tau}{2})^2+
  x^2}$ with $\tau=0.04(1+z)$ (short dashed lines in Figure
\ref{masserr}). This parametrisation of the stellar mass uncertainties
produces too many galaxies with differences larger than 0.2
dex. Finally, we adopt a function which is the product between a
Lorentzian distribution with $\tau=0.04(1+z)$ and a gaussian
distribution with $\sigma=0.5$. This distribution is shown with the
long dashed lines in Figure \ref{masserr}) and provides the best
representation of the stellar mass uncertainties.

\begin{figure}
\centering \includegraphics[width=9cm]{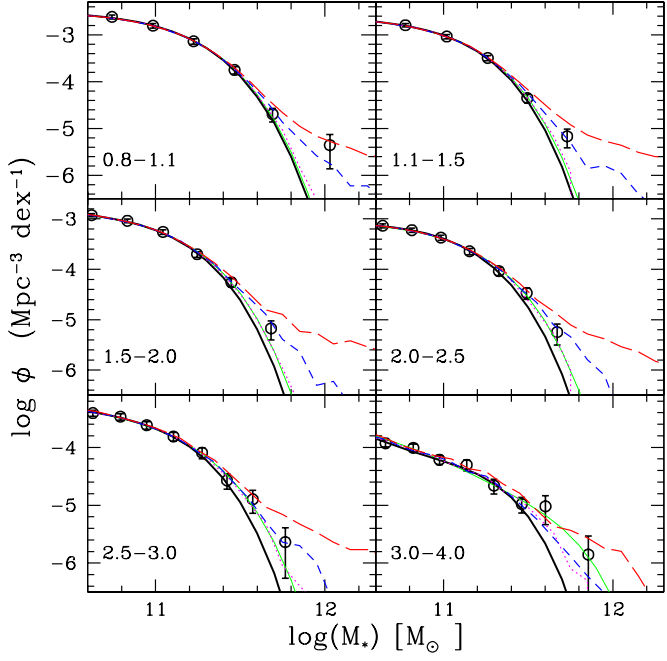}
\caption{Global stellar mass function in several redshift bins. The open circles correspond to the  {\it Vmax} non-parametric estimates. The thick black solid lines correspond to the double-Schechter fit deconcolved by the stellar mass uncertainties. The green thin solid lines  correspond to the simple fit without deconvolution. The pink dotted lines, the short-dashed blue lines and the long dashed red lines correspond to the fiducial mass function (black solid lines) convolved with $G(\sigma=0.04(1+z))$, $L(\tau=0.04(1+z)) G(\sigma=0.5)$, $L(\tau=0.04(1+z)) G(\sigma=1)$, respectively.}
           \label{edd}
\end{figure}

The second step is to take into account the stellar mass uncertainties when
we derive the stellar mass function. We deconvolve only the parametric
fit of the stellar mass function by the stellar mass uncertainties
(not the non-parametric estimates). As described in \S\ref{estimate},
we fit a double-Schechter function $\phi({\cal M})$ to the  {\it Vmax}
non-parametric data. In order to take into account the stellar mass
uncertainties, we convolve the double Schechter function by our
modelized stellar mass uncertainties
\begin{equation}
 \phi_{convolved}({\cal M}) = \int^{\infty}_{-\infty}   \phi(x) L({\cal M}-x,\tau) G({\cal M}-x,\sigma)dx
\end{equation}
with $\tau=0.04(1+z)$ and $\sigma=0.5$. We fit the convolved double
Schechter function to the  {\it Vmax} non parametric data.  Figure \ref{edd}
shows the result of this convolution at $z>0.8$ for the global
sample. The thick solid black lines show the best fit double-Schechter
function, deconvolved by the stellar mass uncertainties (our fiducial
measurement used in this paper). The difference between the
best-fit double-Schechter functions with and without the deconvolution
(black and green lines, respectively) is significant only at the
high-mass end and remains below 0.1 dex. The blue solid dashed lines
correspond to the fiducial MF convolved by the $G\times L$
functions. By construction, the blue dashed line provides a good fit
of the  {\it Vmax} non parametric data points. The pink dotted lines
correspond to the fiducial MF convolved by a simple gaussian with
$\sigma=0.04(1+z)$. As seen in the $0.8<z<1.1$ redshift bin, a
convolution by a gaussian does not explain the presence of some
massive galaxies, which is well explained if we use the $G\times L$
convolution functions. If we allow more extreme errors in the stellar
mass estimate, using for instance a convolution by the same Lorentzian
function multiplied by a Gaussian with $\sigma=1$, we would observe a
larger population of massive galaxies (red long dashed
line). 

\end{document}